\documentclass{article}

\usepackage{changepage}
\usepackage{arxiv}
\usepackage{algorithm}  \usepackage[right]{lineno}
\usepackage{wrapfig}
\usepackage[pscoord]{eso-pic}
\usepackage[fulladjust]{marginnote}
\usepackage{algpseudocode}
{}
{}
{}
\usepackage{xcolor}
\usepackage[normalem]{ulem}
\usepackage{mathrsfs}
\usepackage{relsize}
\usepackage{color}
\usepackage[T1]{fontenc}
\usepackage{amsmath}
\usepackage{amssymb}
 
\usepackage{amsthm} 
\usepackage{mathtools,cuted}
\usepackage{mathrsfs}
\usepackage{bbm}
\usepackage{cite}
 
\usepackage{dsfont}
\usepackage{relsize}
\usepackage{subdepth}
\usepackage{url}

\DeclareFontFamily{OT1}{pzc}{}
\DeclareFontShape{OT1}{pzc}{m}{it}{<-> s * [1.200] pzcmi7t}{}
\DeclareMathAlphabet{\mathpzc}{OT1}{pzc}{m}{it}
\usepackage{filecontents}
\usepackage{stfloats}
\renewcommand\footnotemark{}

\usepackage{bigints}
\usepackage{scalerel}

\usepackage[framemethod=tikz]{mdframed}
\newmdenv[
  hidealllines=true,
  backgroundcolor=blue!10,
  innerleftmargin=8pt,
  innerrightmargin=8pt,
  innertopmargin=0pt,
  innerbottommargin=6pt,
  leftmargin=-0pt,
  rightmargin=-0pt
]{shadedbox}

\definecolor{airforceblue}{rgb}{0.36, 0.54, 0.66}
\definecolor{ballblue}{rgb}{0.13, 0.67, 0.8}

\definecolor{alizarin}{rgb}{0.82, 0.1, 0.26}
\definecolor{asparagus}{rgb}{0.53, 0.66, 0.42}
\definecolor{applegreen}{rgb}{0.55, 0.71, 0.0}
\definecolor{armygreen}{rgb}{0.29, 0.33, 0.13}
\definecolor{amber(sae/ece)}{rgb}{1.0, 0.49, 0.0}
\definecolor{coquelicot}{rgb}{1.0, 0.22, 0.0}
\definecolor{ao(english)}{rgb}{0.0, 0.5, 0.0}
\definecolor{amber}{rgb}{1, 0.6, 0}

\makeatletter
\let\NAT@parse\undefined
\makeatother
\usepackage{hyperref}
\hypersetup{
    colorlinks=true,
    linkcolor=red,
    filecolor=magenta,      
    urlcolor=cyan,
    citecolor=ballblue
}
\usepackage{anyfontsize}
\usepackage{cleveref}
\renewcommand{\hat}{\widehat}

\newcommand{\x}{\boldsymbol{x}}
\newcommand{\R}{\boldsymbol{R}}

\newcommand{\y}{\boldsymbol{y}}
\renewcommand{\d}{\boldsymbol{\delta}}
\renewcommand{\a}{\boldsymbol{\alpha}}

\newcommand{\Y}{\boldsymbol{Y}}
\newcommand{\oo}{\boldsymbol{\Omega}}
\newcommand{\zita}{\boldsymbol{\zeta}}
\newcommand{\e}{\boldsymbol{\epsilon}}
\newcommand{\h}{\boldsymbol{h}}
\newcommand{\uu}{\boldsymbol{u}}
\newcommand{\f}{\boldsymbol{f}}
\newcommand{\g}{\boldsymbol{\gamma}}
\usepackage{arxiv}

\usepackage[utf8]{inputenc} % allow utf-8 input
\usepackage[T1]{fontenc}    % use 8-bit T1 fonts
\usepackage{hyperref}       % hyperlinks
\usepackage{url}            % simple URL typesetting
\usepackage{booktabs}       % professional-quality tables
\usepackage{amsfonts}       % blackboard math symbols
\usepackage{nicefrac}       % compact symbols for 1/2, etc.
\usepackage{microtype}      % microtypography
\usepackage{lipsum}

\title{A Critical Comparizon on Attitude Estimation: From Gaussian Approximate Filters to Coordinate-Free Dual Optimal Control}
\author{
  Nikolas P.~Koumpis$^{1,\dagger}$\thanks{1~\texttt{nikolaos.p.koumpis@gmail.com}} %(nikolaskoumpis.com)}
  \\
  Faculty of Mechanical Maritime and Materials Engineering
  \\Delft Center for Systems and Control \\
  TU Delft\\
  Mekelweg 2
   2628 CD Delft \\
   \And
 Panagiotis A.~Panagiotou$^{2}$\thanks{2~\texttt{p.panagiotou@sheffield.ac.uk}} \\
  Faculty of Electronic and Electrical Engineering\\
  The University of Sheffield\\
  3 Solly Street, Sheffield S1 4DE\\
  \And
 Ioannis.~Arvanitakis$^{3}$\thanks{3~\texttt{ac7632@coventry.ac.uk}} \\
  Faculty of Engineering Environment and Computing\\
  Coventry University\\
   3 Gulson Rd, Coventry CV1 2JH, UK \\

}

\begin{document}
\maketitle

\begin{abstract}
This paper conveys attitude and rate estimation without rate sensors by performing a critical comparison, validated by extensive simulations. The two dominant approaches to facilitate attitude estimation are based on stochastic and set-membership reasoning. The first one mostly utilizes the commonly known Gaussian-approximate filters, namely the EKF and UKF. Although more conservative, the latter seems to be more promising as it considers the inherent geometric characteristics of the underline compact state space and accounts  -from first principles- for large model errors. We address the set-theoretic approach from a control point of view, and we show that it can overcome reported deficiencies of the Bayesian architectures related to this problem, leading to coordinate-free optimal filters. Lastly, as an example, we derive a modified predictive filter on the tangent bundle of the special orthogonal group $\mathbb{TSO}(3)$.
\end{abstract}

% keywords can be removed

\section{Introduction}\label{Section1}
Attitude and rate estimation is an important aspect of aerial robotics. 
Throughout the decades, it has proven very accurate and versatile in applications from the first Low Earth Orbit (LEO) satellites \cite{Maral91_ijsc} to Unmanned Aerial Vehicles (UAVs) \cite{Bouabdallah07_phd} and from the Unmanned Aerial Systems \cite{Gonzalez17_drones} to recent Aerial Robotic Workers \cite{Alexis17_arxiv}. At the same time,  technological and technical advances allow for increased specifications of autonomy in conjunction with precise and agile maneuvering. Consequently, position and orientation (attitude) control constitutes a field of research that is vital component of  aerial robotics.
In many cases, the model can be decoupled and attitude control can
be implemented independently from position control \cite{Bouabdallah07_phd}. Lately, more focus has been given to attitude controllers due to the increased difficulty and complexity of the specific control problem \cite{ANT11_cep}; the success of these controllers relies upon the accurate knowledge of the real orientation and the angular rate of the aerial robot. Thus, it is imperative to develop efficient attitude filters, to deal not only with the measurement noise but also with the model errors.\par

When a-priori statistical information is available, such uncertainties are represented by utilization of the stochastic framework. Subsequently, model errors and measurement noise are then expressed as stochastic inputs to provide a faithful representation of the conditions where the real system operates. Within this probabilistic context, the Bayesian formulation of estimation appears in the form of Gaussian approximate filters. In particular, the Extended Kalman Filter (EKF) \cite{jazwinski2007stochastic, gelb1974applied} and Unscented Kalman Filter (UKF) \cite{julier2000new} constitute traditionally used tools for the problem of attitude and rate estimation, as it appears in aerial robotics. \par
From a series of novel works in the existing literature \cite{lorenz2019attitude,crassidis2003unscented,markley2003attitude, inoue2016extended,firoozi2012analysis}, it is evident how the Gaussian approximate solutions interact with the space of orientations through the various attitude coordinate systems \cite{stuelpnagel1964parametrization, ty1991wen,shuster1993survey, shaub}.
A very fundamental one, being presented in \cite{farrell1970attitude}, expresses the motion using the Euler angles. To avoid the well-known singularity issues, a temporary shifted reference frame is established that estimates the orientation angles w.r.t. the previous angle estimates. By doing so, the representation remains away from singular points. Although the resulted state space model is highly nonlinear due to the involved trigonometric functions, the EKF is used for estimation. The EKF accounts for some drawbacks, especially for highly nonlinear systems. For many applications, derivation of the Jacobian matrices is hard or time consuming. Furthermore, linearization results in an unstable filter performance when the time step intervals for the update are not sufficiently small \cite{hajiyev2017review}. On the contrary, small time steps increase the computational load, especially when the Jacobian matrices are not available in closed form.
\par
Other works address the problem differently by establishing a quasi-linear kinematic expression \cite{markley2004attitude, leffens1982kalman, crassidis2003unscented, chang2016iterated,hua2014robust}. The only attitude representation for this purpose is the algebra of unit quaternions \cite{markley2004attitude}, which is closed under the quaternion multiplication. This nonsingular, four-parameter representation has been discussed by many authors including \cite{leffens1982kalman}. 
Nonetheless, the fact that the correction step of the EKF updates the predicted quaternion by addition results in a corrected (upper part) state that does not express an orientation. For this problem, three solutions exist.  The first one proposes a Euclidean normalization after the correction step; the second one deploys a pseudo-measurement equation; and, finally, the third one is a multiplicative approach proposed by \cite{leffens1982kalman}. The latter is based on the product of the quaternion error and the reference quaternion, both having unit magnitude.

\par

Alternatively, the Unscented Kalman Filter (UKF) has the advantage of handling nonlinearities through the Unscented Transform (UT) more efficiently compared to the EKF.  This makes reasonable the choice for using it in conjunction with the Euler angles coordinate system and the shifted frame of reference method of \cite{farrell1970attitude}. An attempt towards this direction can be found in \cite{cilden2017nanosatellite}. On the other hand, when the quaternion representation is used, the UKF in a standard format cannot be implemented straightforwardly. The reason is again the quaternion's unit constraint. There is no guarantee that the predicted quaternion mean of the UKF will satisfy this constraint and express an orientation. In \cite{crassidis2003unscented}, the authors tackle this obstacle by the use of the Generalized Rodrigues Parameters (GRP) \cite{schaub1996stereographic}
 to represent an attitude-error quaternion. Lastly, a comparison between the EKF and the UKF under the quaternion representation can be found in \cite{laviola2003comparison}. The conclusion is that the UKF shows better performance  compared with the EKF, when the kurtosis and the higher order moments in the state error distributions are significant. A compelling  discussion on the application of the mentioned Kalman-based filters for gyro-less attitude and rate estimation can be found in \cite{hajiyev2017review}. The EKF and UKF are  local methods and are  characterized by relatively small computational complexity. However, they are strictly suboptimal and, thus, they at most constitute efficient heuristics, but
without explicit theoretical guarantees \cite{kalogerias2016grid}.
 \par 
 
An attempt to set the state estimation problem within the dual optimal control framework \cite{kalman1960contributions} was made in \cite{mook1988minimum}. This method determines the corrections added to the assumed model, such that the model and corrections yield an accurate representation of the system's behavior. The model uncertainty is considered as an unknown but deterministic signal within a Hilbert space. The goal is to estimate the states for the resulting measurements to approximate the measured observations, while keeping the considered model as valid as possible. This is done by minimizing the total norm of the augmented measurement-model uncertainty vector. The optimization problem incorporates a covariance constraint in order to ensure that the state estimates remain statistically consistent. However, the above filter is based on a two-point boundary condition problem and is, essentially, an offline optimal state estimator.
In \cite{Mortensen}, the modal trajectory estimator is derived. This approach is based entirely on the Hamiltonian formulation of optimal control and results in a recursive filter.
\par
The importance of the dual optimal control formulation for the problem of attitude and rate estimation stems from the nature of orientation itself.  Euler's theorem \cite{euler} indicates that the set of orientations is the special orthogonal group $\mathbb{SO}(3)$, which is a compact Lie group associated with the Lie algebra $\mathfrak{so}(3)$ of the $3\times3$ skew symmetric matrices. A Lie group is a differentiable manifold equipped with the algebraic structure of a group \cite{stramigioli2001geometry,galier}. Therefore, instead of relying on the prefabricated Bayesian architectures, the problem can be directly set and solved in a coordinate-free fashion as a dual optimal control problem by applying tools from differential geometry.
The approach of \cite{Mortensen}, commonly known as minimum energy filtering, was utilised in \cite{saccon} where the second-order-optimal minimum energy filter on Lie groups was derived.
\par
Conclusively, we observe that within the -Bayesian framework- the success of a gyroless attitude estimation scheme depends primarily on the chosen coordinate system. Essentially, there is an incompatibility between the Bayesian architectures and the space of orientation. This incompatibility is justified by the fact that the Gaussian approximate filters are primarily built  to approximate the conditional mean, rather than comply with the geometric characteristics of the underlying state-space.
 \par 
Furthermore, based on stochastic modelling, the Bayesian strategies assume second-order statistical knowledge for both the measurement noise and the model's uncertainty. Although aggregating second order statistics for the measurement noise is feasible through (offline) experimentation, for the case of the model error -usually referred to as "process noise"- the assumption that it is a symmetrically-distributed white noise process of known covariance has no theoretical basis. For physical systems, model uncertainty represents environmental phenomena; therefore, it is more reasonably expressed by smooth functions within a Hilbert space.
 \par

\par
With regards to the dual optimal control formulation, deterministic filtering originates from set membership estimation, where the prior and the underlying uncertainties are expressed as assigned -from the modeler- sets. On the one hand, although intuitively the set-membership reasoning seems compatible with the compact nature of the space of orientations, it does not provide any accuracy about the belief degree. On the other hand, dual optimal control provides the machinery to formulate the estimation problem as a well-defined optimization problem \cite{zhang2011attitude}.

 \par
In this paper, we consider sensors that measure only vector directions and we incorporate both the kinematic and the dynamic models for the attitude motion. 
The set of observations, are made w.r.t. the inertial frame and obtained from sensors that measure w.r.t. the body frame.
The contribution of this work is the critical assessment of the reasons governing the superiority of deterministic modelling over stochastic,  for the problem of orientation and rate estimation from vector measurements. Although many works study various attitude filters' performances in terms of attitude and rate error accuracy, none of them is motivated by the fact that deterministic modelling naturally leads to a coordinate-free problem formulation. To this extent, the present paper is motivated by the dual optimal control approach, that accounts directly both for the underlying state-space and the environmental phenomena affecting the existing system without ad-hoc simplification assumptions. To this direction, we also derive the modified predictive filter on  $\mathbb{TSO}(3)$. Extensive simulations are used to compare the second-order-optimal minimum energy filter (MEF) \cite{saccon} and predictive filter (PF) performance versus the EKF and UKF. Both the analysis and the simulations' results conclusively indicate that coordinate-free deterministic filtering tackles the vices of the stochastic approach.
\par

\textbf{Bayesian formulation of attitude estimation and Gaussian approximate filters (Section~\ref{Section2}).} After establishing a coordinate system map for the orientation we end up with a system of nonlinear differential equations. We may then attack the problem under stochastic reasoning and in particular through the Bayesian approach. We present the Bayesian formulation of attitude estimation and employ the Kushner equation \cite{kushner1964differential} for analysis.\\ \textbf{Set membership state estimation and dual optimal control formulation (Section~\ref{Section3}).} A more conservative point of view considers set-membership reasoning for the problem at hand. We show that set-membership estimation approach naturally leads to a control formulation of estimation, which is optimally implemented by the minimum energy filter. In the same section, we derive the modified predictive filter on $\mathbb{TSO}(3)$ by proposing a new error function.\\ \textbf{Algorithms and Numerical Implementation (Section~\ref{Section4})}. We provide the algorithm summaries for each filter providing exceptional care to the numerical integration. On the one hand, explicit integration methods add, artificially, energy into the system. On the other hand, implicit integration schemes operate as (artificial) dampers removing energy from the system. To this direction,
we utilize the Lie group symplectic integration, which essentially is produced under the machinery of the variational principle of mechanics by discretizing directly the cost functional \cite{marsden}.\\ \textbf{Simulation Results (Section~\ref{Section5}).} We present the results for two case studies regarding UAVs and two for satellite attitude filtering. In particular, we demonstrate how the filters operate under the presence of process noise and significant deterministic model errors.\\ \textbf{Simulation Results (Section~\ref{Section6}).} The paper concludes with remarks that are drawn based on the obtained results.\par
\textit{Notation}: The following notation is used throughout the paper: $\mathbb{R}$ is the set of real numbers. With $\text{rod}: q\rightarrow R$ we declare the Rodrigues formula which maps the quaternion $q$ (or the principal rotation vector) to the Directional Cosine Matrix (DCM) $R \in \mathbb{SO}(3)$. The matrix $\text{expm(X)}$ is the exponential of $X \in \mathbb{R}^{n\times n}$. The map $(~ )^\times$:$ \mathbb{R}^3\rightarrow \mathfrak{so}(3)$ is an isomorphism from the arrays in $\mathbb{R}^3$ to the Lie algebra of the $3\times 3$ skew symmetric matrices $\mathfrak{so}(3)$. The Euclidean norm is denoted by $||~||$. $\nabla{V}$ is the gradient of the real-valued function $V: \mathbb{R}^{n}\rightarrow \mathbb{R}$ and $\partial_{\boldsymbol{X}}(f)$ denotes the partial derivative of $f$ w.r.t. $\boldsymbol{X}$. Further, $\langle \boldsymbol{X} ,\boldsymbol{Y} \rangle : \mathbb{R}^n\times\mathbb{R}^n\rightarrow \mathbb{R}$ denotes the inner product $\forall ~\boldsymbol{X}, \boldsymbol{Y} \in \mathbb{R}^n$. Lastly, the estimate of $\boldsymbol{X}$ is denoted by $\widehat{\boldsymbol{X}}$, while the optimal estimate of $X$ by $\widehat{\boldsymbol{X}}^*$.
\section{Bayesian formulation of attitude estimation and Gaussian approximate filters}\label{Section2}

Let $({\Omega}, {{F}}, {\mathcal{P}})$ be the filtered probability space and the filtration $\mathbb{F}_{t}$ with respect to which all processes will be adapted. After establishing a  coordinate system map, we consider the following processes of interest: 
\begin{equation}\label{eq1}
\begin{aligned}
{d\boldsymbol{X}_{1}}&={\boldsymbol{f}_1}({\boldsymbol{X_1}},\boldsymbol{X}_{2}) dt~,\\
{d \boldsymbol{X}_{2}}&={\boldsymbol{f}_2}({\boldsymbol{X_{2}}},\boldsymbol{u}) dt+{{G^{\frac{1}{2}}}} {d}\boldsymbol{W}~,~
\end{aligned}
\end{equation}
\begin{equation}\label{eq2}
\hspace{-0.7cm}d\boldsymbol{Y}=\boldsymbol{h}(\boldsymbol{X_1}, t)dt+d\boldsymbol{V}~,~
\end{equation}
where the state process $\boldsymbol{X}=\big\{ \boldsymbol{X}_t=[\boldsymbol{X}^{\top}_{1,t}~~\boldsymbol{X}^{\top}_{2,t}]^{\top},~t\geq0 \big\}$ is defined to be the solution of the stochastic differential equation \eqref{eq1} and equation \eqref{eq2} defines the observation process $\boldsymbol{Y}=\big\{\boldsymbol{Y}_t, t\geq0\big\}$. Furthermore, $\boldsymbol{W} \in \mathbb{R}^{n_2}$ and $\boldsymbol{V} \in \mathbb{R}^{q}$ express environmental effects and the measurement noise respectively and are assumed to be independent Brownian motions. The coefficients $\boldsymbol{f}=[\boldsymbol{f}^{\top}_{1}~~\boldsymbol{f}^{\top}_{2}]^{\top}: \mathbb{R}^{{n_1}+{n_2}}\rightarrow \mathbb{R}^{{n_1}+{n_2}}$ and
$\boldsymbol{h}:{\mathbb{R}}^{n_1}\rightarrow {\mathbb{R}}^{q}$ 
are assumed to be Lipschitz continuous mappings. Lastly, the control $\boldsymbol{u} \in \mathcal{U}  \subseteq \mathbb{R}^{n_2} $ is considered as known input torques and $G^{\frac{1}{2}}$ is the square root of $G \in \mathbb{R}^{n_2\times n_2}$.
The coefficients $\boldsymbol{f}_{1}$ and $\boldsymbol{f}_{2}$ express the kinematics and the dynamics of the physical motion respectively, as indicated by the Euler's equations of motion \cite{shaub}. The process $\boldsymbol{X}$ comprises the orientation and angular rate respectively, where $n_1$ depends on the chosen coordinate system map. 
%%The system operates in the real world; thus, it is influenced by uncertain effects, expressed by $\boldsymbol{W}$.
%As the state process evolves continuously, sensors measure nonlinear functions of the system's state according to 
%\begin{equation}
%y=h(x, t)+\epsilon_{t}~,~
%\end{equation}
%with $h:\mathbb{R}^n \times T \rightarrow \mathbb{R}^q$. The observations are corrupted by Gaussian measurement noise $\epsilon\in \mathbb{R}^q$, independent of $w$, with mean zero and covariance $\mathbb{E}\{\epsilon(t)\epsilon^T(\tau)\}=Q(t)\delta_{t-\tau}~ \forall t,\tau$.
%The functions $f$ and $h$, the prior information regarding the state, along with the noise distributions, are sufficient in order to formulate the problem under the Bayesian approach.
\par 
\vspace{0.1cm}
We denote by $\mathscr{F}^{\mathscr{Y}}_t$ the $\sigma$-algebra generated by $\{\boldsymbol{Y}_\tau,~0\leq \tau \leq t\}$. The optimal  estimate  $\widehat{\boldsymbol{X}}_t^*$ is then given as the solution to the following optimization problem:
\begin{equation}\label{eq3}
\begin{aligned}
{\widehat{\boldsymbol{X}}_t^*}=\text{arg}\min_{{\widehat{\boldsymbol{x}}}}\int_{\mathbb{R}^n}\mathcal{{C}}({\boldsymbol{x}},{\widehat{\boldsymbol{x}}})\rho(\boldsymbol{x}|\mathscr{F}^{\mathscr{Y}}_t)d{\boldsymbol{x}}~,~
\end{aligned}
\end{equation}
where $\rho(\boldsymbol{x}|\mathscr{F}^{\mathscr{Y}}_t)$ is the conditional probability density of the state, given the noisy measurements up to and including time $t$. Therefore, knowledge of the posterior density for each $t$, constitutes the complete solution of the problem \eqref{eq3}. 
For ${C}({\boldsymbol{x}},{\widehat{\boldsymbol{x}}})= ||\boldsymbol{x}-\widehat{\boldsymbol{x}}||^{2}_2$, the optimal Mean Square Error (MSE) estimate is given by
\begin{equation}\label{eq4}
\begin{aligned}
{\widehat{\boldsymbol{X}}_t^*}= \mathbb{E}\{\boldsymbol{X}_t|\mathscr{F}^{\mathscr{Y}}_t \} \equiv \int_{\mathbb{R}^n}{\boldsymbol{x}}\rho(\boldsymbol{x}|\mathscr{F}^{\mathscr{Y}}_t)d\boldsymbol{x}
\end{aligned}
\end{equation}
In order to derive a differential equation for the optimal MSE estimate, we can differentiate \eqref{eq4} w.r.t. time. By utilizing the generalized Leibniz rule  \cite{flanders1973differentiation} we obtain:
\begin{equation}\label{eq5}
d{\widehat{\boldsymbol{X}}^*}=\int_{\mathbb{R}^n}{\boldsymbol{x}}\frac{\partial \rho(\boldsymbol{x}|\mathscr{F}^{\mathscr{Y}}_t)}{\partial t}d{\boldsymbol{x}}
\end{equation}
Furthermore, the posterior density $\rho=\rho(\boldsymbol{x}|\mathscr{F}^{\mathscr{Y}}_t)$ evolves according to the Kushner equation \cite{kushner1964differential}:
\begin{equation}
\begin{aligned}
\frac{\partial{\rho}}{\partial t}\hspace{-0.1cm} &=\rho(d\boldsymbol{Y}-\mathbb{E}\{\boldsymbol{h}|\mathscr{F}^{\mathscr{Y}}_t\}dt)^{\top}{}(\boldsymbol{h}-\mathbb{E}\{ \boldsymbol{h}|\mathscr{F}^{\mathscr{Y}}_t\})
-\sum_{k=1}^{n}\frac{\partial}{\partial x_k}\left({\boldsymbol{f}_k} \rho \right)+\frac{1}{2}\sum_{k,l=1}^{n}\frac{\partial}{\partial x_k\partial x_{_l}}({G}_{k,_{l}}\rho)~.
\end{aligned}
\end{equation} 
Thus, the optimal nonlinear filter is given by:
\begin{equation}\label{eq7}
\begin{aligned}
{\widehat{\boldsymbol{X}}}_{k,t}^*&= \int^{t}_{0}\mathbb{E}\{ \boldsymbol{f}_k|\mathscr{F}^{\mathscr{Y}}_s\}ds
+\int^{t}_{0}\boldsymbol{\mathcal{V}}_s^{\top}\big(d\boldsymbol{Y}_s-\mathbb{E}\{\boldsymbol{h} |\mathscr{F}^{\mathscr{Y}}_s\}
ds\big)
\end{aligned}
\end{equation}
with $k=1,...,n_1+n_2$~ and $\boldsymbol{\mathcal{V}}_{k,s}$ satisfying the stochastic differential equation:
\begin{equation}\label{eq7}
\begin{aligned}
d\boldsymbol{\mathcal{V}}_{k,t}=d\big( \mathbb{E}\{ \boldsymbol{h}_{k}{x}_k|\mathscr{F}^{\mathscr{Y}}_s\}-\mathbb{E}\{\boldsymbol{h}_{k} |\mathscr{F}^{\mathscr{Y}}_s\}\mathbb{E}\{ {X}_k|\mathscr{F}^{\mathscr{Y}}_s\}\big)
\end{aligned}
\end{equation}
Nevertheless, actual evaluations of the terms $\mathbb{E}\{ \boldsymbol{f}_k|\mathscr{F}^{\mathscr{Y}}_s\}$, $\mathbb{E}\{ \boldsymbol{h}_k|\mathscr{F}^{\mathscr{Y}}_s\}$ are possible only in case where the system is linear and the noise distributions are Gaussian, resulting in the well known Kalman filter \cite{kalman}. In the nonlinear case, both terms require knowledge of the entire posterior density, yielding an infinite dimensional filter \cite{maybeck1982stochastic}. The EKF and the UKF are proposed to tackle this issue. The EKF applies the Kalman filter framework to nonlinear systems, by first linearizing the system model using a first-order truncated Taylor series expansion around the current estimates \cite{haykin}. 
This linearization step affects the accuracy of the posterior predictions and often leads to divergence of the filter \cite{haykin}. On the contrary, the UKF \cite{julier1995new} makes explicit use of the scaled Unscented Transformation (UT) (stochastic linearization) \cite{julier2002scaled} which is based on the idea that, it is preferable to approximate a probability distribution instead of an arbitrary nonlinear function. Since both filters utilize only the Gaussian parameters (first and second statistical moments), these methods belong to a broader class entitled as Gaussian approximate filters. 
\par
Implementing both the EKF and UKF with the quaternion representation yields bilinear kinematic equations. However, there is no guarantee that the quaternion mean of the EKF and UKF will satisfy the unit-norm constraint due to the addition operator in the correction step of the filters. To overcome this issue, in this work we use the Euclidean normalization approach \cite{bar1971optimum}. Although this normalization step leads to meaningful results, it is an external intervention on both the EKF and UKF algorithms and affects the unbiasedness of both the quaternion and rate estimates. This phenomenon is discussed and analysed further in the \textbf{(Appendix \ref{AppendixC})}. 

\section{Set membership state estimation and dual optimal control formulation}\label{Section3}

Orientation belongs in a compact space, the special orthogonal group. An assumption that is valid in many applications is that the angular rate lies within a bounded space. The same can be inferred for both the model and measurement uncertainty. Thus, instead of modeling uncertainties utilizing stochastic reasoning, we use the more elementary concept of a set \cite{talak2019theory}. This section shows how deterministic filtering naturally recasts as a control problem adopting set-theoretic reasoning.
 
Consider the system described by the state-space model of the form:
\begin{equation}\label{eq9}
\begin{aligned}
{\dot{\boldsymbol{x}}_{1}}&={\boldsymbol{f}_1}({\boldsymbol{x}_{1}},\boldsymbol{x}_{2})\\
{\dot{\boldsymbol{x}}_{2}}&={\boldsymbol{f}_2}(\boldsymbol{x}_{2},\boldsymbol{u})+{G}{\boldsymbol{\delta}}~,~
\end{aligned}
\end{equation}
Without loss of generality, $t_i\in I$, where $I$ is a partition of time. Thus, the measurement equation is given by
 \begin{equation}\label{eq10}
\boldsymbol{y}={\boldsymbol{h}}(\boldsymbol{x}_1, t_i)+\boldsymbol{\epsilon}~,~
\end{equation}
where the functions $\boldsymbol{f}_k$, $k=1,2$, $G$ and $\boldsymbol{h}$ are defined as in Section \ref{Section2}.
Regarding \eqref{eq9}, \eqref{eq10}, the model and measurement uncertainties are considered as unknown and deterministic signals where $\boldsymbol{\delta} \in \mathcal{D}$ and  $\boldsymbol{\epsilon} \in \mathcal{E}$, with $\mathcal{D}\subset \mathbb{R}^{n_2}$ , $\mathcal{E} \subset \mathbb{R}^6$. The system's state $\boldsymbol{x} \in \mathcal{S}\times \mathcal{W}$, where $S$ declares the space of orientation and $\mathcal{W} \subset \mathbb{R}^{n_2}$ a properly chosen $C\text{-set}$ \cite{blanchini2008set}. Assuming complete lack of knowledge regarding the initial state estimate, we can write $\widehat{\mathcal{X}}_0 = \mathcal{S}\times \mathcal{W}$. Set-membership state estimation repeats the following two steps \cite{lavalle2012sensing}: 

The guess $\widehat{\mathcal{X}}_{i}$ regarding the state $\boldsymbol{x}$ at time $t_i$ is projected forward in time, resulting the set 
\begin{equation}\label{eq7}
\mathscr{R}_{\mathcal{D},{\boldsymbol{u}_i}} = \{ \boldsymbol{\sigma}|\boldsymbol{\sigma}=\boldsymbol{f}(\boldsymbol{x}_i,\boldsymbol{u}_i)+\boldsymbol{\delta}_i , ~\boldsymbol{x}_i \in \widehat{\mathcal{X}}_i,~ \boldsymbol{\delta}_i\in\mathcal{D}\}
\end{equation}
of all reachable states at time $t_{i+1}$ given $\uu_i$, for all $\d_i \in \mathcal{D}$. Subsequently, at $t_{i+1}$, ~$\mathscr{R}_{\mathcal{D},{\boldsymbol{u}_i}} $ is refined to 
\begin{equation}\label{eq8}
\mathscr{C}_{\mathcal{E},{\boldsymbol{y}_i}}=\{\boldsymbol{x} | \boldsymbol{y}_{i+1}=\boldsymbol{h}(\boldsymbol{x}_{1,i+1},t_{i+1})+\boldsymbol{\epsilon}_{i+1}\}~,~
\end{equation}
which consists of all the states in $\mathcal{S}\times \mathcal{W}$ compatible with the measurement $\boldsymbol{y}_{i+1}$ for some $\boldsymbol{\epsilon}_{i+1} \in \mathcal{E}$. 
The prediction and correction of the state are then given as:
\begin{equation}\label{eq87}
\widehat{\mathcal{X}}_{i+1|i}=\mathscr{R}_{\mathcal{D},{\boldsymbol{u}_i}} ~,~
\end{equation}
and 
\begin{equation}\label{eq88}
\widehat{\mathcal{X}}_{i+1|i+1}=\widehat{\mathcal{X}}_{i+1|i} ~\mathlarger{\cap}~ \mathscr{C}_{\mathcal{E},{\boldsymbol{y}_{i+1}}}~,
\end{equation}
respectively. 
Note that \eqref{eq87} and \eqref{eq88} correspond to the prediction and correction step in the optimal Bayesian update \cite{maybeck1982stochastic} respectively. However, in this case we can go a step further by defining the input pair $(\boldsymbol{x}_0,\boldsymbol{\delta}_{[0,i]})$ and  write, $\boldsymbol{h}(\boldsymbol{x}_{i+1},t_{i+1})=\boldsymbol{h}(\phi(\boldsymbol{x}_0,\boldsymbol{u}_{[0,i]},\boldsymbol{\delta}_{[0,i]}),t_{i+1})$, where $\phi$ is the solution of \eqref{eq9}  and $\boldsymbol{\delta}_{[0,i]}$, $\boldsymbol{u}_{[0,i]}$ declare the model error and input values respectively within $[t_0,t_i]$. Then, the second step of the method is equivalently  modified by defining the set 
\begin{equation}\label{eq8}
 \begin{aligned}
\mathscr{D}_{\mathcal{E},{\boldsymbol{y}_{i+1}}}\hspace{-0.7mm}=\hspace{-0.5mm}\{(\boldsymbol{x}_{0,i},\boldsymbol{\delta}_{[0,i]}) | \boldsymbol{h}(\phi(\boldsymbol{x}_{0,i},\boldsymbol{u}_{[0,i]},\boldsymbol{\delta}_{[0,i]}),t_{i+1})+\hspace{-0.4mm}\boldsymbol{\epsilon}_{i+1}\hspace{-0.6mm}=\hspace{-0.6mm}\boldsymbol{y}_{i+1}\hspace{-0.4mm}\}
\end{aligned}
\end{equation}
of all input pairs which produce observations compatible with the measurements. In other words,  the goal is to actually determine the set of different decisions $\boldsymbol{\delta}_{[0,i]}$ and the initial state $\widehat{\mathcal{X}}_{0,i}$ that produce -throughout the dynamics- the received measurements for $\boldsymbol{\epsilon}_{i} \in \mathcal{E}$. 
Lastly, it is possible to ask for the pair $(\widehat{\mathcal{X}}_{0,i},\boldsymbol{\delta}_{[0,i]})$ such that
 \begin{equation}\label{eq82}
 \begin{aligned}
\mathscr{M}_{\mathcal{E},{\boldsymbol{y}_{i+1}}}\hspace{-0.7mm}=\hspace{-0.5mm}\{(\widehat{\mathcal{X}}_{0,i},\boldsymbol{\delta}_{[0,i]}) | \mathlarger{\mathlarger{\mathlarger{\wedge}}}_{_{k=1}}^{i}\boldsymbol{h}(\phi(\boldsymbol{x}_{0,k},\boldsymbol{u}_{[0,k]},\boldsymbol{\delta}_{[0,k]}),t_{k+1})+
\boldsymbol{\epsilon}_{k+1}\hspace{-0.6mm}=\hspace{-0.6mm}\boldsymbol{y}_{k+1}, \boldsymbol{\epsilon}_{k+1} \in \mathcal{E}, \d_{[0,k]} \in \mathcal{D}^{k} \hspace{-0.4mm}\}
\end{aligned}
\end{equation}
and expect $\widehat{\mathcal{X}}_{0,i}  \downarrow \widehat{\mathcal{X}}^*_0$, i.e. $\widehat{\mathcal{X}}_{0,i}$ to be a decreasing sequence with limit the optimal estimate $\widehat{\mathcal{X}}^*_0$. 
Nonetheless, the set-theoretic algorithm in its general form accounts for some difficulties: The performance of the above method depends on the initial guess $\widehat{\mathcal{X}}_0$, as well as on our knowledge regarding the sets $\mathcal{D}$ and $\mathcal{E}$. Subsequently, representing the sets $\mathcal{X}_0$, $\mathcal{D}$, $\mathcal{E}$, $\mathscr{R}_{\mathcal{D},{\boldsymbol{u}_i}} $ and $\mathscr{C}_{\mathcal{E},{\boldsymbol{y}_{i+1}}}$ $(\mathscr{D}_{\mathcal{E},{\boldsymbol{y}_{i+1}}})$ in practical applications -at least approximately- by a finite set of parameters, is not a trivial task \cite{hanebeck2001recursive}. Finally, the method does not provide any accuracy about the belief degree regarding the state estimates.
\subsection{Minimum energy filtering}\label{Section3a}

The set-theoretic approach determines possible sequences of decisions. The minimum energy filter considers the sequence with the minimum norm which creates observations compatible with the obtained measurements, and constitutes one of the first implementations of this approach. It was first introduced by Mortensen \cite{Mortensen}, and consists of a method for deriving nonlinear estimators, based on the value function of the optimal estimation problem.

Consider the system described by \eqref{eq9}. The signals $\d(\cdot)$, $\e(\cdot)$ and the initial condition $x_0$ are now modeled as arbitrary disturbances within a Hilbert space. Thus, consider the cost 
\begin{equation}\label{eq13}
J(\d_{[t_0,t]},\e_{[t_0,t]}, \x_0;t) =S_{0}\left(\x_{0}\right)+\frac{1}{2} \int_{t_{0}}^{t}\hspace{-0.2cm} \Phi(\d)+Q(\e) d \tau~,~
\end{equation}
where $\d_{[t_0,t]}$ and $\e_{[t_0,t]}$ refer to the model and measurement error values within the interval $[t_0, t]$. Furthermore, $\Phi : \mathbb{R}^{n_2} \rightarrow \mathbb{R}_{+}$ and $Q : \mathbb{R}^6 \rightarrow \mathbb{R}_{+}$ are two quadratic forms that measure the instantaneous energy of the error signals. In addition, $S_0 : \mathbb{R}^{n_1+n_2} \rightarrow \mathbb{R}_{+}$ is the initial cost encapsulating the a-priori knowledge regarding the state at time $t_0$ and is a function with a global minimum \cite{saccon}. Since $\e$ is deterministic, \eqref{eq13} can be written as 
\begin{equation}\label{eq14}
J(\d_{[t_0,t]}, \x_0;t)\hspace{-0.1cm} =\hspace{-0.1cm}S_{0}\left(\x_{0}\right)+\frac{1}{2} \int_{t_{0}}^{t}\hspace{-0.2cm} \Phi(\d)+Q(\y-{\h}({\x_1}, \tau)) d \tau~.
\end{equation}
Note that in order for the filter to track the actual measurements, the uncertainties $\d(\cdot)$ and $S_0(\x_0)$ should be minimal; within the estimation context, minimizing $\int_{t_{0}}^{t} \Phi(\d)d\tau$ is essential rather than an additional requirement as it is posed in classic optimal control theory. The minimization of the uncertainty regarding the actual system is equivalent to the information gain. It is impossible to track the actual system or equivalently estimate the system's state without minimizing the uncertainty for the actual system. Therefore, the goal is to minimize the model uncertainty, while tracking the given measurements. This will yield an optimal minimum energy pair $(\x^*_0, \d^*_{[t_0, t]})$, with the end point of the optimal trajectory $\widehat{\x}^*_{[t_0,t]}=\phi(x^*_0, \d^*_{[t_0, t]},\uu_{[t_0,t]})$ being the minimum energy state estimate $\widehat{\x}^*_{[t_0,t]}(t)$ at time $t$. Thus, the following optimization problem 
\begin{equation}\label{eq15}
\begin{aligned}
\min_{\d_{[t_0, t]},\x_0} \quad & J(\d_{[t_0,t]}, \x_0;t) \\
\textrm{s.t.} \quad & {\dot{\x}_{1}}={\f_1}({\x_{1}},\x_{2})\\
&{\dot{\x}_{2}}={\f_2}(\x_{2},\uu)+{G}{\d}~,~
\end{aligned}
\end{equation}
has to be solved for each $t$ as new observations arrive online, since the optimal decisions $\d^*_{[t_0,t]}$ are affected from the incoming information at each time instant $t$. At this point, we follow \cite{omar} where \eqref{eq15} is tackled by first assuming fixed $\x_0$ and finding the optimal $\d^*_{[t_0,t]}$ with the Hamiltonian formulation of optimal control providing the necessary conditions for optimality for $\d^*_{[t_0,t]}$ \cite{athans2013optimal}. The value function is defined as 
\begin{equation}\label{eq16}
V(\x_{[t_0,t]};t)=\min_{\d_{[t_0, t]}}  J(\d^*_{[t_0,t]}, \x_0;t)~. \\
\end{equation}

In order to completely solve the optimization problem of \eqref{eq15}, the minimum of the value function w.r.t. the initial condition $\x_0$ for each $t$ must be considered. The necessary condition for optimality yields:
\begin{equation}\label{eq18}
\nabla V(\x_{[t_0,t]};t)_{\x_{t_0}}=0~\forall t~,
\end{equation}
However, \eqref{eq18}  is equivalent with 
\begin{equation}\label{eq19}
\nabla V(\x_{[t_0,t]};t)_{\x^*_{t}}=0~\forall t~,
\end{equation}
since determining the optimal end point $\x^*_{t}$-and given the optimal control decisions $\d^*_{[t_0,t]}$- fully specifies the optimal initial condition $\x_0$ for each $t$, by running time backwards. Essentially, this equivalence allows us to express the value function w.r.t. the optimal estimate $\widehat{\x}^*_t$ and, therefore, to derive the minimum energy filter \hspace{-0.1mm}\cite{Mortensen}.
\subsection{Predictive filter on $\mathbb{TSO}(3)$}\label{Section3b}

The predictive filter on $\mathbb{TSO}(3)$ is a deterministic filter that predicts the model error and drives the rate and attitude estimate towards the real state under the presence of significant model errors. The filter emerges from the continuous-time nonlinear controller of \cite{chen2003optimal} along with the covariant constraint from \cite{doi:10.2514/3.20302}. A predictive quaternion attitude filter based on the nonlinear controller of \cite{lu1994nonlinear} was derived in \cite{crassidis1997predictive}. However, our derivation is based on a different cost function which leads to a faster transient response. Furthermore, the output Jacobians are determined intrinsically, directly on the $\mathbb{TSO}(3)$ as shown in the \textbf{(Appendix~\ref{AppendixA})}. 

Consider the state space model:
\begin{equation}\label{eq36}
\begin{aligned} 
\dot{R}(t) &=R(t)\oo^{\times}(t) \\
 \dot{\oo} (t)&=\mathbb{I}^{-1}\left((\mathbb{I}\oo(t))^{\times} \oo(t)+\boldsymbol{T}(t)\right)+G\d(t)\\ 
 \y(t)&=\left[\begin{array}{l}{\y_{1}(t)} \\ {\y_{2}(t)}\end{array}\right]=\left[\begin{array}{c}{R(t)^{\top} \a_{1}(t)} \\ {R(t)^{\top} \a_{2}(t)}\end{array}\right]+D \e(t) ~,~
 \end{aligned}
\end{equation}
where $R \in \mathbb{SO}(3)$, $\oo \in \mathbb{R}^{3}$, $\a_i:\mathbb{R}_{+} \rightarrow \mathbb{R}^3$. The vector $\e$ represents the unknown measurement error with $D$ being block diagonal, namely 
\begin{equation}\label{opt}
D=\left[\begin{array}{cc}
d_{1} I_{3 \times 3} & 0 \\
0 & d_{2} I_{3 \times 3}
\end{array}\right].
\end{equation}

Given that the term $\widehat{\y}_i^{\times}\y_i$ forms an error axis between the estimated output $\widehat{\y}_i$ and the system's output $\y_i$, it is reasonable to target for the model error that minimises the predicted mean error axis formed by the two measurements. Based on this observation, the predictive filter on $\mathbb{TSO}(3)$ results from the following minimization problem:
\begin{equation}\label{eq37}
\begin{aligned}
\min_{\d(t)} \quad & \frac{1}{2}\bigg\|\sum_{i=1,2}{{\widehat{\y}_i}^{\times}(t+h)\y_i(t+h)}\bigg\|_{Q^{}}^2
%^{T} Q(y(t+h)-\hat{y}(t+h))\\
+\frac{1}{2}||\d(t)||_{\Sigma}^2~,%^T(t)\Sigma \delta(t)
\\
\textrm{s.t.} \quad &\dot{\widehat{R}}(t) = \widehat{R}(t)\widehat{\oo}^{\times}(t) \\ 
& \dot{{\widehat{\oo}}}(t) = \mathbb{I}^{-1}((\mathbb{I}{\widehat{\oo}}(t) )^{\times} {\widehat{\oo}}(t)+\boldsymbol{T}(t)) +{G}\d(t) \\ 
& \widehat{\y}(t)=\left[ \begin{array}{c} {\widehat{\y}_{1}(t)}\\ {\widehat{\y}_{2}(t)}\end{array}\right] =\left[ \begin{array}{c} {{\widehat{R}}^{T}(t) {{\a}}_{1}(t)} \\ {{\widehat{R}}^{T}(t) {{\a}}_{2}(t)}\end{array}\right]~,
\end{aligned}
\end{equation}
where the matrices $Q \in \mathbb{R}^{3\times3}$ and $\Sigma \in \mathbb{R}^{3\times3}$ penalise the prediction error and the correction term  respectively.
The value of the uncertainty term $\d$ at time $t$ influences the state $(R,\oo)$ at a posterior instant of time 
$t+h$ and, subsequently, the same is true for the output since the state-output relation is expressed via a memoryless system.
The constrained optimization problem of \eqref{eq37} recasts into an unconstrained one by using the expansion \cite{brockett1976functional,gilbert1977functional}:
\begin{equation}\label{eq40}
 \widehat{\y}_k(t+h) \approx \widehat{\y}_k(t)+{\zita_i}(\widehat{R},\widehat{\oo}, h;t)+{\Lambda}(h) \mathbb{W}_k(\widehat{R},\widehat{\oo})\d(t)~,
\end{equation}
where 
\begin{equation}\label{eq41}
\zita_{k}(\widehat{R},\widehat{\oo},h;t)= h\mathcal{L}_{\f}^{1}(\widehat{\y}_k)+\frac{\h^2}{2!}\mathcal{L}_{\f}^{2}(\widehat{\y}_k)~,
\end{equation}
and
\begin{equation}\label{eq42} 
\Lambda(h)=\frac{h^2}{2}\boldsymbol{I}_{3\times3}~.
\end{equation}
Term $\mathcal{L}_{\f}^{\xi}(\widehat{\y}_i), \xi=1,2$ denotes the $\text{$\xi$-th}$ order Lie derivative of $\widehat{\y}_i$ w.r.t.\ the system. After substituting \eqref{eq40} in the cost, the necessary condition for optimality yields the optimal correction term:
\begin{equation}
 \d^{*}(t) = -\frac{1}{2}\hspace{-0.1cm}\left( \mathcal B^{\top}Q^{-\top}\mathcal B + \Sigma^{\top}\right)^{-1}\hspace{-0.15cm} \cdot \mathcal B^{\top}\hspace{-0.1cm}\hspace{-0.1cm}\cdot \big( Q^{-1} +Q^{-\top} \big) \g(t)~,
 \label{eq48}
\end{equation}
where $\mathcal B$ is a function of $\widehat{R},\widehat{\oo}$, and $h$, given by:
\begin{equation}
\mathcal B(\widehat{R},\widehat{\oo},h)=\sum_{k=1,2} \y^{\times}_k\Lambda(\h)w_k(\widehat{R},\widehat{\oo})~,
\end{equation}
$\g(t)$ is given by:
\begin{equation}
\g(t)=\sum_{i=1,2}\y_k^{\times}\widehat{\y}_k^{}+\sum_{k=1,2}\y_k^{\times}\zita_{k}(\widehat{R},\widehat{\oo},h;t)~,
\end{equation}
and $w_k$ is given by:
\begin{equation}
w_k(\widehat{R},\widehat{\oo})= (\widehat{R}^{\top}\a_k)^{\times} {G}~.
\end{equation}
Lastly, by substituting the Lie derivative terms (Appendix \ref{AppendixA}), equation \eqref{eq41} results in
\small
\begin{equation}
\zita_k(\widehat{R},\widehat{\oo},h;t) \hspace{-0.1cm} = \hspace{-0.1cm} \frac{h^2}{2}\Big\{\hspace{-0.08cm}(\widehat{R}^{}\a_k)^{\times}{\mathbb{I}^{-1}\hspace{-0.1cm} \left(\hspace{-0.05cm}(\widehat{\oo} \mathbb{I})^{\times} \widehat{\oo}+\boldsymbol{T}\right)}\hspace{-0.1cm}+\hspace{-0.1cm}(\widehat{\oo}^{\times})^{2}\widehat{R}^{\top}\a_k\Big\}~.
\end{equation}
The main advantage of this method is that the correction is performed only through the dynamics, while the kinematic equation remains isolated; consequently, it can be integrated geometrically. This was not the case in the Gaussian approximate filters, where the addition operator in the correction step violates the space's geometry. Furthermore, there is no need to initialise the filter with prior information.
 
Until now, the problem has been treated as a tracking problem of optimal control. However, the estimates should be statistically consistent. As can be seen from \eqref{eq48}, by decreasing the model error penalty matrix $\Sigma$, the estimates are based more on the measurements, so the output estimates get closer to the noisy observations. Assuming white measurement noise, a limit must be set w.r.t. how much the estimated outputs should match the noisy observations. 
This is accomplished by choosing the model error penalty matrix $\Sigma$ such that it approximately achieves the balance expressed by
\begin{equation}\label{constr}
\mathbb{E}\left\{(\widehat{\y}(t)-\Y_t))(\widehat{\y}(t)-\Y_t)^{\top}\right\}\approx\sigma_{\epsilon} I_{6 \times 6}~,
\end{equation}
referred as the covariant constraint \cite{doi:10.2514/3.20302}.
For our application, we estimate the output error covariance by
\begin{equation}\label{s3:eq:outcov}
M=\frac{1}{N}\sum^N_{k} (\widehat{\y}_k-\y_k)(\widehat{\y}_k-\y_k)^{\top}~,
\end{equation}
where $N$ is the total number of samples. To examine \eqref{constr}, we utilise the $\text{L}_{2,2}$ matrix norm. Since the measurement noise covariance matrix is of the form $D=\sigma I_{6\times6}$, \eqref{constr} is satisfied when: 
\begin{equation}
\sigma^{*}=\arg\min_{\sigma}\Big(\sum_{k,j}(M_{k,j}-D_{k,k})^2\Big)^{\frac{1}{2}}~,
\end{equation}
which after some calculations, yields

\begin{equation}\label{eq36}
\sigma^{*}=\frac{\text{trace}(M)}{6}~.
\end{equation}

\section{Algorithms and Numerical Implementation}\label{Section4}

An extensive simulation study is carried out to compare the performance of the second-order-optimal MEF and PF against the EKF and UKF.
In this section, the model that is utilised in the simulations and the error functions used to assess the efficiency of the methods are presented. Also the algorithmic summaries are given for each of these four filters and some aspects relating to the numerical implementation are presented.

The EKF and UKF use the quaternion representation, whereas the MEF and PF are set directly on the special orthogonal group. Although many works study the performance of various attitude filters in terms of attitude and rate error accuracy, none of them does so by considering dynamics with significant model errors. Attitude and rate estimation from vector measurements should take into account environmental phenomena which affect the actual system. In \cite{dos2009attitude}, the $4$-\text{th} order Runge-Kutta method is employed for simulation. Nevertheless, these methods do not preserve the continuous-time motion's essential features like kinetic energy and momentum. The main contributions to address these gaps are:
Algorithm summaries for each of the aforementioned attitude filters, as well as 
a comprehensive simulation study that compares the selected stochastic attitude filters against the deterministic ones. The comparison considers measurement errors, initialization errors, and model errors that typically appear in attitude and angular rate filtering for UAVs and satellite missions. 

\subsection{Model and Error function}\label{Section4a}
For expressing the orientation of the rigid body, we use the quaternion representation $q \in \mathbb{S}^{3}$ and the matrix representation $R \in \mathbb{SO}(3)$. 
Then, the rigid body kinematics are given as:
\vspace{-0.2cm}
\begin{equation}
\dot{q}(t)=\frac{1}{2}M(\oo(t))q(t)~,
\end{equation}
\vspace{-0.1cm}
and 
\vspace{-0.1cm}
\begin{equation}
\dot{R}(t)=R(t)\oo^{\times}(t)~,
\end{equation}
where 
\begin{equation}
M(\oo) =\frac{1}{2}\left[\begin{array}{cc}0 & -\oo^{\top} \\ \oo & -\oo^\times\end{array}\right].
\end{equation}
Expressed in the body-fixed frame, we denote by $\mathbb{I} \in \mathbb{R}^{3\times3}$ the inertia tensor, by $\oo \in \mathbb{R}^3$ the angular rate of the rigid body and by $\boldsymbol{T} \in \mathbb{R}^3$ the applied torques.
The angular rate $\oo$ evolves according to Euler's equation \cite{marsden}:
\vspace{-0.2cm}
\begin{equation}
\dot{\oo} (t)=\mathbb{I}^{-1}\left((\mathbb{I}\oo(t))^{\times} \oo(t)+\boldsymbol{T}(t)\right)+G\d(t)\end{equation}
up to model uncertainty $\d(t) \in \mathbb{R}^3$, with $G \in \mathbb{R}^{3\times3}$. 
Two time varying directions $\a_1(t)$ and $\a_2(t)$ are measured on board, as $\y_1(t)$ and $\y_2(t)$ according to: 

\begin{equation}
 \y(t)=\left[\begin{array}{l}{\y_{1}(t)} \\ {\y_{2}(t)}\end{array}\right]=\left[\begin{array}{c}{r(q(t))^{\top} \a_{1}(t)} \\ {r(q(t))^{\top} \a_{2}(t)}\end{array}\right]+D \e(t) ~,~
\end{equation}

\begin{equation}
 \y(t)=\left[\begin{array}{l}{\y_{1}(t)} \\ {\y_{2}(t)}\end{array}\right]=\left[\begin{array}{c}{R(t)^{\top} \a_{1}(t)} \\ {R(t)^{\top} \a_{2}(t)}\end{array}\right]+D \e(t) ~,~
\end{equation}
where $r \in \mathbb{SO}(3)$ is the Directional Cosine Matrix (D.C.M.) parameterised w.r.t the unit quaternion ${q}(t)$, and $\e(t)$ is the measurement noise. We assume that the two sensors operate independently, so the matrix $D$ is chosen block diagonal:
\begin{equation}
D=\left[\begin{array}{cc}
d_{1} I_{3 \times 3} & 0 \\
0 & d_{2} I_{3 \times 3}
\end{array}\right]
\end{equation}

The attitude estimation error is given by the following functions:
 \begin{equation}
\begin{aligned} 
e_{q}(t) &= \cos^{-1} \bigg(  1- \frac{\text{tr}( I - r(q(t))^{\top}{r(\widehat{q}(t))})}{2}    \bigg)
\end{aligned}
\end{equation}
and
\begin{equation}
\begin{aligned} 
e_{R}(t) &= \cos^{-1} \bigg(  1- \frac{\text{tr}( I - R^{\top}(t){\widehat{R}(t))})}{2}    \bigg)
\end{aligned}
\end{equation}
w.r.t. the quaternion and matrix representation, respectively. 
The angular rate estimation error is calculated as $\boldsymbol{e}_{\oo}=\widehat{\oo}(t)-\oo(t)$ where both $\oo(t)$ and $\widehat{\oo}(t)$ are expressed w.r.t. the inertial frame. 
\par

\subsection{Numerical implementation}\label{Section4b}
In this section, discrete-time implementations of the continuous-time filters are presented via algorithm summaries. Discretization should be addressed carefully as the Lie group structure of the underlying state space; the motion's energy and momentum have to be preserved under any numerical calculation. Proper discretization of the continues-time differential equations requires Lie group variational (symplectic) integration \cite{marsden}.
The numerical integration of the kinematic equation is made by assuming a short-time step $h$. Since the attitude motion is instantaneously a rotation, the discrete orientation update is obtained using the exponential map as:
\begin{equation}\label{eq38}
q_{k+1}=\frac{1}{2} \exp\left(hM\left(\oo_{k}\right)\right) q_k,
\end{equation}
w.r.t. the quaternion representation, and as 
\begin{equation}\label{eq39}
R_{k+1}=R_{k} \exp \left(h {\oo_{k}}^{\times}\right)
\end{equation}
in terms of $R \in \mathbb{SO}(3)$.
The angular velocity update emerges by employing a Newton solver for 
\begin{equation}
C_{\exp }\left(-h \oo_{k+1}\right)\left(\mathbb{I} \oo_{k+1}\right)=
  C_{\exp }\left(h \oo_{k}\right)\left(\mathbb{I} \oo_{k}\right)+h \boldsymbol{U}_k~,
  \end{equation}
where $\boldsymbol{U}$ is the control vector and
\begin{equation}
C_{\exp }(X)=\mathbb{I}_{3 \times 3}-\frac{1}{2} X+\frac{1}{12}\left(X^{\times}\right)^{2}~.
\end{equation}
In this work, the physical motion, the prediction step of the EKF, and state propagation of the sigma points in the UKF are made using the Lie group symplectic integration \cite{bou2009hamilton, Kobilarov:2009:LGI,bottasso2008multibody}.
\par 
An important flaw related with the EKF's and UKF's  implementation is the singularity of the state estimation error covariance matrix, when the orientation is expressed by the unit quaternions. The unit norm constraint results in the singularity of the latter covariance matrix \cite{leffens1982kalman}. Three solutions to this problem exist \cite{leffens1982kalman}. In this work, regarding the UKF's implementation, we utilise the approach that deletes one of the quaternion components in order to obtain a truncated state error covariance expression. Per contra, because the EKF consists of second order terms only, it does not compute an ill-conditioned covariance matrix. This claim is mathematically justified in \cite{carmi2007covariance}.
\begin{algorithm}\small
\caption{EKF for attitude and rate estimation}
\begin{algorithmic}[1]   
\State $\hat{x}_{0|0}=[\hat{q}_0; \hat{\Omega}_0]$, $P_{0|0}=P_0$
\\ 
\State \textbf{for k=1,2,...} \\
\State ~~~~~\textbf{Solve for} $\oo_{k+1}$ \textbf{using a Newton solver} 
\State ~~~~~$C_{\exp }\left(-h \oo_{k+1}\right)\left(\mathbb{I} \oo_{k+1}\right)=
C_{\exp }\left(h \oo_{k}\right)\left(\mathbb{I} \oo_{k}\right)+h U_k$
  ~\\
\State ~~~~~\textbf{Update} $\widehat{q}_{k}$ \textbf{using Euler's theorem:}
\State ~~~~~$\widehat{q}_{k+1}=M(\oo_{k})\widehat{q}_k$
   \\
\State ~~~~~$\widehat{\x}_{k+1|k}=[\widehat{q}_{k+1}^{\top}, \widehat{\oo}^{\top}_{k+1}]^{\top}$   
   \\
\State ~~~~~$P_{k+1|k}=F(\widehat{\x}_{k+1|k},u_{k})P_{k|k}F^T(\widehat{\x}_{k+1|k},\uu_{k})+W$  
       \\                             
\State ~~~~~$\y_{i,k+1|k}=r(q_{k+1})\a_{i,k},~i=1,2$
    \\ 
\State ~~~~~$P^{\y}_{k+1|k} =H(\x_{k+1})P_{k|k+1}H(\x_{k+1})^{\top}+Q$
\\ 
\State ~~~~~$P^{\x\y}_{k+1|k} = P_{k+1|k}H(\x_{k+1})^T $
\\
\State ~~~~~$K_{k+1}=P^{\x \y}_{k+1|k}(P^{\y}_{k+1|k})^{-1}$
\\
\State ~~~~~$\hat{\x}_{k+1|k+1}=\hat{\x}_{k+1|k}+K_{k+1}(\y_{k+1}-\hat{\y}_{k+1|k})$
\\
\State ~~~~~$P_{k+1|k+1}=P_{k+1|k}-K_{k+1} P^{\y}_{k+1|k} K_{k+1}^{\top}$
\\
\State ~~~~~$\widehat{\x}_{{k+1|k+1},[1:4]}=\widehat{\x}_{{k+1|k+1},[1:4]}~{||~\widehat{\x}_{{k+1|k+1},[1:4]}~||_2}^{-1}$
\\
\State \textbf{end for}
\end{algorithmic}
\end{algorithm}
\begin{algorithm}\small
\caption{UKF for attitude and rate estimation}
\begin{algorithmic}[1] 
\State $\widehat{\x}_{0|0}=[\widehat{q}_0; \widehat{\oo}_0]$, $P_{0|0}=P_0$ 
\\
\State \textbf{for k=1,2,...}\\
\State ~~~~~$V_k=P_{k}+R$\\
\State ~~~~~\textbf{Calculate sigma points }$\boldsymbol{S}_{k}$ \textbf{based on} $(\x_k,C(V_k))$\\
\State ~~~~~\textbf{Time update:}
\\
\State ~~~~~${\boldsymbol{S}}_{k+1 \mid k}={f}\left({\boldsymbol{S}}_{k}, {\uu}_{k}\right)$
\\
\State ~~~~~$\widehat{{\x}}_{k+1|k}=\sum_{i=0}^{2L} w_{i}^{(m)} {\boldsymbol{S}}_{i, k+1 \mid k}$
\\
\\
$\hspace{-0.5mm}{P}_{{\x}_{k+1|k}}\hspace{-0.8mm}=\hspace{-0.5mm}\sum_{i=0}^{2 L}\hspace{-0.5mm} w_{i}^{(c)}\hspace{-1mm} \left({\boldsymbol{S}}_{i, k+1 \mid k}-\widehat{{\x}}_{k+1|k}\right)\hspace{-1.5mm}\left({\boldsymbol{S}}_{i, k+1 \mid k}-\widehat{{\x}}_{k+1|k}\hspace{-1mm}\right)^{\top}$
\\
\State ~~~~~\textbf{Calculate output prediction sigma points: }
\\
\State ~~~~~${\Y}^p_{i,k+1 \mid k}={\boldsymbol{H}}\left({\boldsymbol{S}}_{i,k+1 \mid k}^{\x},t\right)$\\
    \algstore{myalg}
\end{algorithmic}
\end{algorithm}
\begin{algorithm}                     
\begin{algorithmic} [1]                   % 
\algrestore{myalg}
\State ~~~~~\textbf{Average }
\\
~~~~~$\widehat{{\y}}_{k+1|k}=\sum_{i=0}^{2L} w_{i}^{(m)} {\Y}^p_{i, k \mid k-1}$\\

%\algstore{myalg}
%\end{algorithmic}
%\end{algorithm}
%\begin{algorithm}                     
%\begin{algorithmic} [1]                   % enter  
%\algrestore{myalg}  

\State \textbf{Measurement update }
\State \hspace{-1.5mm}$\hspace{-0.5mm}\begin{aligned} 
{P}_{e} &=\sum_{i=0}^{2L} w_{i}^{(c)}\hspace{-0.1cm}\left({\Y}^p_{i, k+1 \mid k}-\widehat{{\y}}_{k+1|k}\right)\hspace{-1.5mm}\left({\Y}^p_{i, k \mid k-1}-\widehat{{\y}}_{k+1|k}\right)^{\top} \hspace{-1.8mm}+ \hspace{-0.9mm}{R} \\ 
{P}_{{x}_{} {y}_{}} &=\sum_{i=0}^{2L} w_{i}^{(c)}\left({\boldsymbol{S}}_{i, k+1 \mid k}-\widehat{{\x}}_{k+1|k}\right)\left({\Y}^p_{i, k+1 \mid k}-\widehat{{\y}}_{k+1|k}\right)^{\top} \\ {K}_{k} &={P}_{{\x}_{} {\y}_{}} {P}_{e}^{-1} \\ 
\widehat{{\x}}_{k} &=\widehat{{\x}}_{k+1|k}+{K}_{k}\left({\y}_{k}-\widehat{{\y}}_{k+1|k}\right) \\ &{P}_{{\x}_{k+1|k+1}} ={P}_{{\x}_{k+1|k}}-{K}_{k} {P}_{e} {K}_{k}^{\top} \end{aligned}$\\
\State ~~~~~$\widehat{\x}_{{k+1|k+1},[1:4]}=\widehat{\x}_{{k+1|k+1},[1:4]}~{||~\widehat{\x}_{{k+1|k+1},[1:4]}~||_2}^{-1}$\\
\State \textbf{end for}
\end{algorithmic}
\end{algorithm}
\newpage
where $C(V_k)$ refers to the square root of $V_k$ resulting from the Cholesky factorization.
 The second-order-optimal minimum energy filter on $\mathbb{TSO}(3)$ is implemented based on the $(0)-$connection function \cite{saccon}.
\begin{algorithm}[h]\small
\caption{MEF for attitude and rate estimation}
\begin{algorithmic}[1] 
\State 
 $K_{0}=I_{6\times6}$, $ \widehat{R}_0=I_{3}$, $\widehat{\oo}=[0\ 0\ 0]^{\top}$
\\
\State \textbf{for k=1,2,...}\\
\State ~~~~~$\boldsymbol{r}_{R}=-\sum_{i=1,2}{{\widehat{\y}_i}^{\times}\y_i}$\\
\State ~~~~~$\widehat{R}(k+1)=\widehat{R}(k)\text{expm}(h(\widehat{\Omega}_k+K_{11}(k)\boldsymbol{r}_R))$\\
\State~~~~~ \textbf{Solve for }$\widehat{\oo}_{k+1}$ \textbf{using a Newton solver:}
\\
\State$C_{\exp }\left(-h \widehat{\oo}_{k+1}\right)\left(\mathbb{I} \widehat{\oo}_{k+1}\right)= C_{\exp }\left(h \widehat{\oo}_{k+1}\right)\left(\mathbb{I} \widehat{\oo}_{k}\right)+hK_{21}(k)r_R$
\\
\\
\State $A=\left[\begin{array}{cc}-\widehat{\oo}^{\times} & I_{3}\\ 0 & \mathbb{I}^{-1}\left[(\mathbb{I} \widehat{\oo})^{\times}-\widehat{\oo}^{\times} \mathbb{I}\right]\end{array}\right]$
\\
\\
\State $E=\left[\begin{array}{cc}\sum_{i=1}^{2}-\left(q_{i} / d_{i}^{2}\right)\left(\widehat{\y}_{i}^{\times} \y_{i}^{\times}+\y_{i}^{\times} \widehat{\y}_{i}^{\times}\right) / 2 & 0 \\ 0 & 0_{3 \times 3}\end{array}\right]$\\
\State $B R^{-1} B^{\top}=\left[\begin{array}{cc}0_{3 \times 3} & 0 \\ 0 & B_{2} R^{-1} B_{2}^{\top}\end{array}\right]$\\
\State $W\left(K, \boldsymbol{r}_{R}\right)=\left[\begin{array}{cc}1 / 2\left(K_{11} \boldsymbol{r}^{R}\right)^{\times} & 0 \\ 0 & 0_{3 \times 3}\end{array}\right]$\\
    \algstore{myalg}
\end{algorithmic}
\end{algorithm}
\begin{algorithm}                     
\begin{algorithmic} [1]                   % 
\algrestore{myalg}
\\
\State \begin{equation*}
\begin{aligned} 
{K}(k+1)=-\alpha K(k)+A K(k)+K(k) A^{\top}\\
-K(k) E K(k)+B R^{-1} B^{\top} \\ 
-W\left(K(k), \boldsymbol{r}_{R}\right) K(k)-K(k) W\left(K(k), \boldsymbol{r}_{R}\right)^{\top} 
\end{aligned}
\end{equation*}
\\
\State \textbf{end for}
\end{algorithmic}
\end{algorithm}
\newpage
where $K_{11}(k)=K_{[{1:3},{1:3}]}(k)$ and $K_{21}(k)=K_{[{4:6},{1:3}]}(k)$

\begin{algorithm}\small
\caption{PF for attitude and rate estimation}
\begin{algorithmic}[1]   
\State $\widehat{\x}_{0}=(\widehat{R}_0, \widehat{\oo}_0)$
\\
 \State \textbf{for k=1,2,...} \\
\State ~~~~~\textbf{Update} $\widehat{R}_{k}$ 
\\
\State ~~~~~$\widehat{R}_{k+1}=\widehat{R}_k\text{expm}(h\widehat{\oo}^{\times}_k)$
\\
\State ~~~~~\textbf{Calculate}~$\zita_i(\widehat{R}_k,\widehat{\oo}_k,h;t)$ 
   \\
\State ~~~~~\textbf{Calculate} $w_i(\widehat{R}_k,\widehat{\oo}_k,h)$ 
   \\
\State ~~~~~\textbf{Calculate}$~~B(\widehat{R},\widehat{\oo},h)$ and $\boldsymbol{\gamma}_k$
\\
\State ~~~~~\textbf{Calculate} $\delta^*_k$
\\
 \State ~~~~~\textbf{solve for }$\widehat{\oo}_{k+1}$
 \textbf{using a Newton solver:}
\State $C_{\exp }\left(-h \oo_{k+1}\right)\left(\mathbb{I} \oo_{k+1}\right)=
C_{\exp }\left(h \oo_{k}\right)\left(\mathbb{I} \oo_{k}\right)+h( \uu_k+\mathbb{I}\d^*_k)$
\\

\State \textbf{end for}
\\
\State \textbf{if} ~~~$\frac{trace(M)}{6} < \sigma_\epsilon$ 
\\
 \State ~~~~~~~~~$\Sigma \downarrow$
 \\
 \State \textbf{else if} ~~~~~~$\frac{trace(M)}{6} > \sigma_{\epsilon}$
 \\
  \State ~~~~~~~~~$ \Sigma \uparrow$
  \\
 \State \textbf{else} 
 \\
 \State ~~~~~~~~~~~~\textbf{keep} $\Sigma$
 \\
 \State \textbf{end if}
\end{algorithmic}
\end{algorithm}

 \section{Simulation Results}\label{Section5}
In this section, we describe a series of simulations for two distinct cases. We demonstrate attitude and rate estimation from vector measurements for UAVs and LEO satellites. For both case studies, the measurement noise and model uncertainty are initially modeled as white Gaussian noises. Subsequently, to stress the significance of the dual optimal control formulation, we replace the model error with an unknown deterministic disturbance that exerts on the existing system.
\subsection{Simulation Cases}\label{Section5a}

\subsubsection{Case 1: Attitude and rate estimation for UAVs}
n this case, the measurement noise is Gaussian zero mean random process and it is set relatively large aiming to express poor sensor quality. In particular, matrix $D$ is chosen so that the signals
$d_i\epsilon_i(t),~i=1,2$ have standard deviations of $20^o$ degrees.
The initial orientation and rate deviation are kept at normal levels and  are initialised with the unit quaternion $q_0=(70^o,[1~1~1]^{\top})$ and the angular rate $\oo_0=[0.3~0.2~0.1]^{\top}~\text{rad/sec}$. The initial orientation matrix is obtained by using the Rodrigues formula $R_0=\text{rod}(q_0)$ \cite{hashim2019special}.
The control torques are given by 
$\boldsymbol{T}(t)=[\sin \left(\frac{2 \pi}{3} t\right) ~-\sin \left(\frac{2 \pi}{1} t\right)  ~\cos \left(\frac{2 \pi}{5} t\right)]^{\top}$. Lastly,
we assume that the two reference vectors $\a_{i}(t),~i=1,2$ are orthogonal for every $t$. Table \ref{table:1} and   \ref{table:2} summarise the system's and filter's parameterization, respectively.

\begin{table}[h!]
\begin{center}
\scalebox{0.8}{
\begin{tabular}{|p{3cm}||p{9cm}|}
 \hline
Time Step & $0.001(s)$ \\
\hline
Simulation Time & $100(s)$ \\
\hline
Initial orientation & $q_0=[0.8253,0.3260,0.3260,0.3260]^{\top}$ ~$R_0=\text{rod}(1.2,[1,1,1])$\\
\hline
Initial rate & $\oo_0=[0.2, 0.4, 0.5]^{\top}$ \\ 
\hline
Inertia tensor & $\text{diag}(6,7,9)$ \\
\hline
Reference directions & $\a_1(t)=[1, 0, 0],~\a_2(t)=[0, 1, 0]$ \\
\hline
Input torque & $[\sin \left(\frac{2 \pi}{3} t\right) -\sin \left(\frac{2 \pi}{1} t\right)  \cos \left(\frac{2 \pi}{5} t\right)]^{\top}$ \\
\hline
Model error (AWGN)& $\mathcal{N}(0,0.1)$\\
\hline
Model error & $0.1*[\sin \left(\frac{2 \pi}{5} t\right) -\sin \left(\frac{2 \pi}{5} t\right)  \cos \left(\frac{2 \pi}{5} t\right)]^{\top}$\\
\hline
Measurement error & $\mathcal{N}\sim (0, 20)$ \\
\hline
\end{tabular}}
\caption{UAV Parameters}
\label{table:1}
\end{center}
\end{table}

\begin{table}[h!]
\begin{center}
\scalebox{0.8}{
\begin{tabular}{ |p{3cm}||p{5cm}|}
\hline
Time Step & $0.001(s)$ \\
\hline
Simulation Time & $40(s)$ \\
\hline
EKF  &$P_0= I_{7\times 7}$   \\
\hline
UKF  &$P_0= I_{7\times 7}$   \\
\hline
MEF  &$K_0=I_{6\times6}$   \\
\hline
PF     &$Q=10^{-3},~\Sigma=5\cdot10^{-3}$ \\
\hline
\end{tabular}}
\end{center}
\caption{Filters' initialisation for UAV's attitude and rate estimation}
\label{table:2}
\end{table}
\subsubsection{Case 2: Attitude and rate estimation for satellite mission}
In this case, we consider smaller measurement noise levels. The input torques are also assumed of lower frequency and the inertia tensor is increased resulting a slow satellite's motion. 
The initial orientation deviates significantly from the identity since the spacecraft can be oriented arbitrarily around its center of mass. The initial angular rate is set smaller compared to the previous experimental study declaring the much slower motion of the satellite. The parameters of the system and the initialization parameters of the filters are summarised in Table \ref{table:3} and Table \ref{table:4}, respectively.
\begin{table}[h!]
\begin{center}
\scalebox{0.8}{
\begin{tabular}{ |p{3cm}||p{9cm}|}
 \hline
Time Step & $0.001(s)$ \\
\hline
Simulation Time & $40(s)$ \\
\hline
Initial orientation & $q_0=[0.4085,0.5270,0.5270,0.5270]^{\top}$ $R_0=\text{rod}(2.3,[1,1,1])$\\
\hline
Initial rate & $\oo_0=[0.1, 0.3, 0.2]^{\top}$ \\ 
\hline
Inertia tensor & $\text{diag}(102,105,103)$ \\
\hline
Reference directions & $\a_1(t)=[1, 0, 0],~\a_2(t)=[0, 1, 0]$ \\
\hline
Input torque & $[\sin \left(\frac{2 \pi}{25} t\right) -\sin \left(\frac{2 \pi}{13} t\right)  \cos \left(\frac{2 \pi}{37} t\right)]^{\top}$ \\
\hline
Model error (AWGN)& $\mathcal{N}(0,0.1)$\\
\hline
Model error & $0.1*[\sin \left(\frac{2 \pi}{13} t\right) -\sin \left(\frac{2 \pi}{12} t\right)  \cos \left(\frac{2 \pi}{17} t\right)]^{\top}$\\
\hline
Measurement error & $\mathcal{N}\sim (0, 20)$ \\
\hline
\end{tabular}}
\caption{Satellite Parameters}
\label{table:3}
\end{center}
\end{table}

\begin{table}[h!]
\begin{center}
\scalebox{0.8}{
\begin{tabular}{ |p{3cm}||p{5cm}|}
 \hline
Time Step & $0.001(s)$ \\
\hline
Simulation Time & $40(s)$ \\
\hline
EKF  &$P_0= I_{7\times 7}$   \\
\hline
UKF  &$P_0= I_{7\times 7}$   \\
\hline
MEF  &$K_0=I_{6\times6}$   \\
\hline
PF   &$Q=10^{-3},~\Sigma=5\cdot10^{-3}$\\
\hline
\end{tabular}}
\end{center}
\caption{Filters' initialisation for satellite attitude and rate estimation}
\label{table:4}
\end{table}
\newpage

\subsection{Results}\label{Section5b}
\subsubsection{Case 1: Attitude and rate estimation for a UAV}\label{Section5a}

\begin{figure}[h!]
\begin{center}
\begin{tabular}{c}
{\includegraphics[scale=0.15]{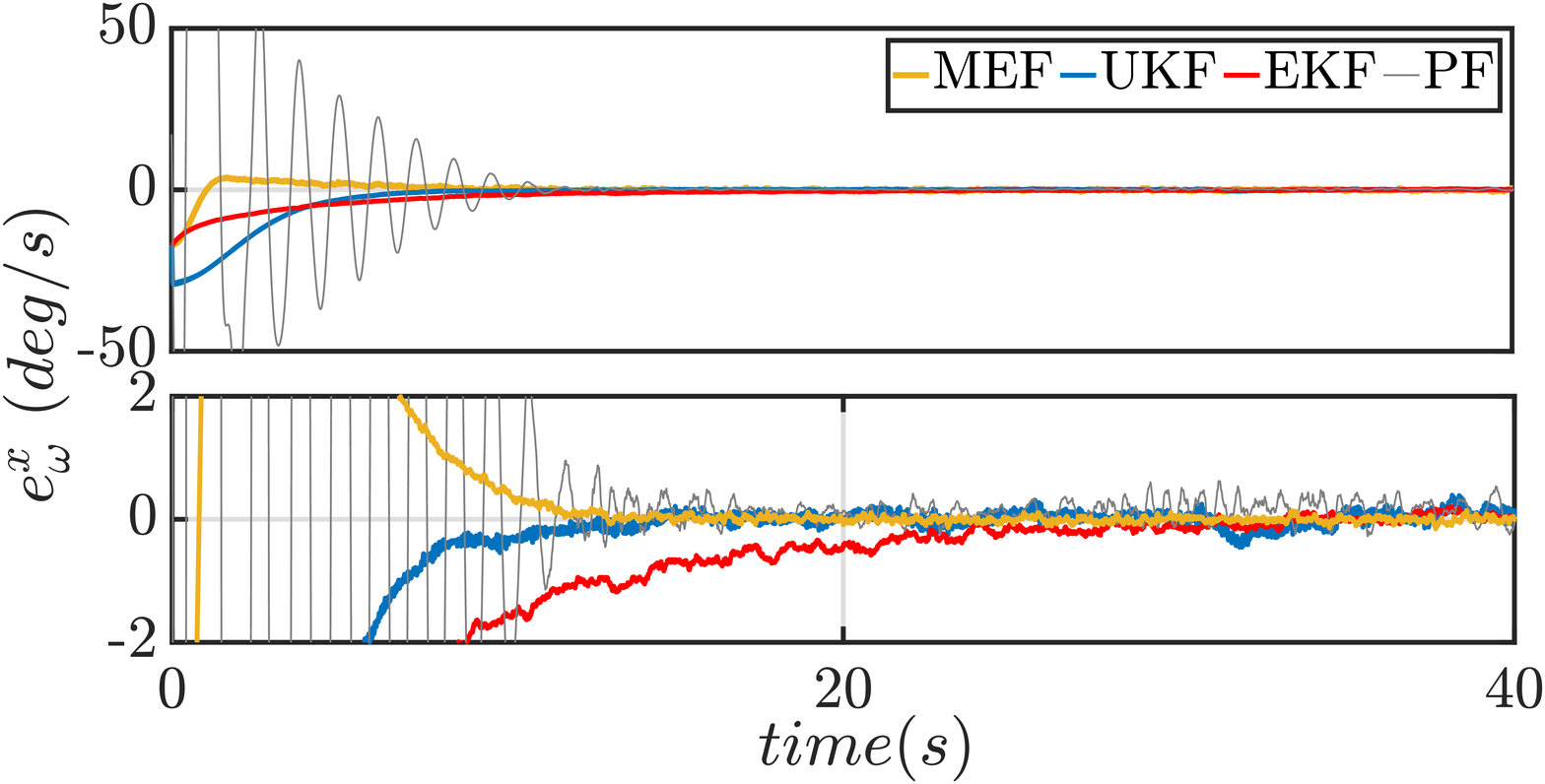}} 
\vspace{-0.4cm}
\\{\includegraphics[scale=0.15]{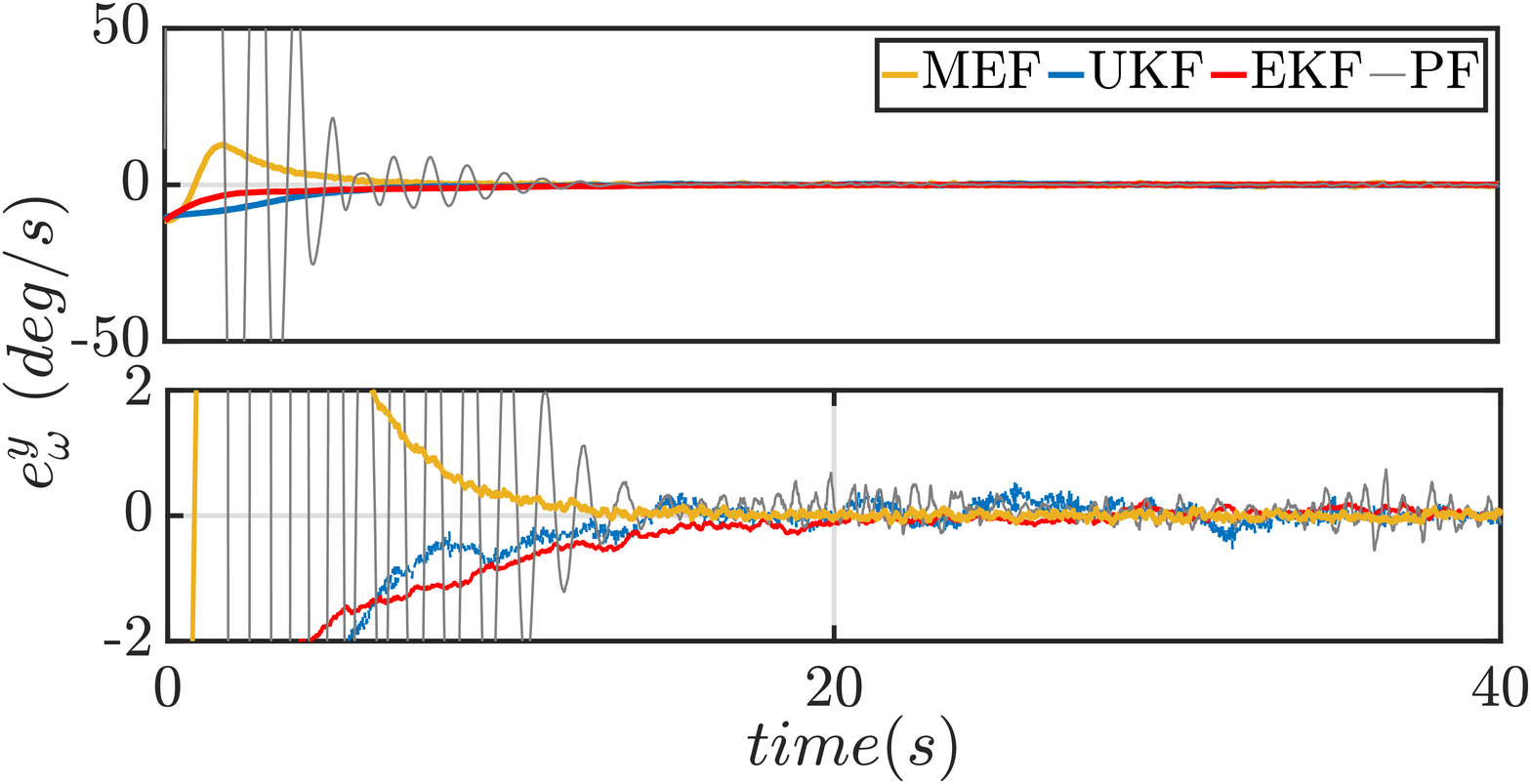}}
\vspace{-0.4cm}
\\{\includegraphics[scale=0.15]{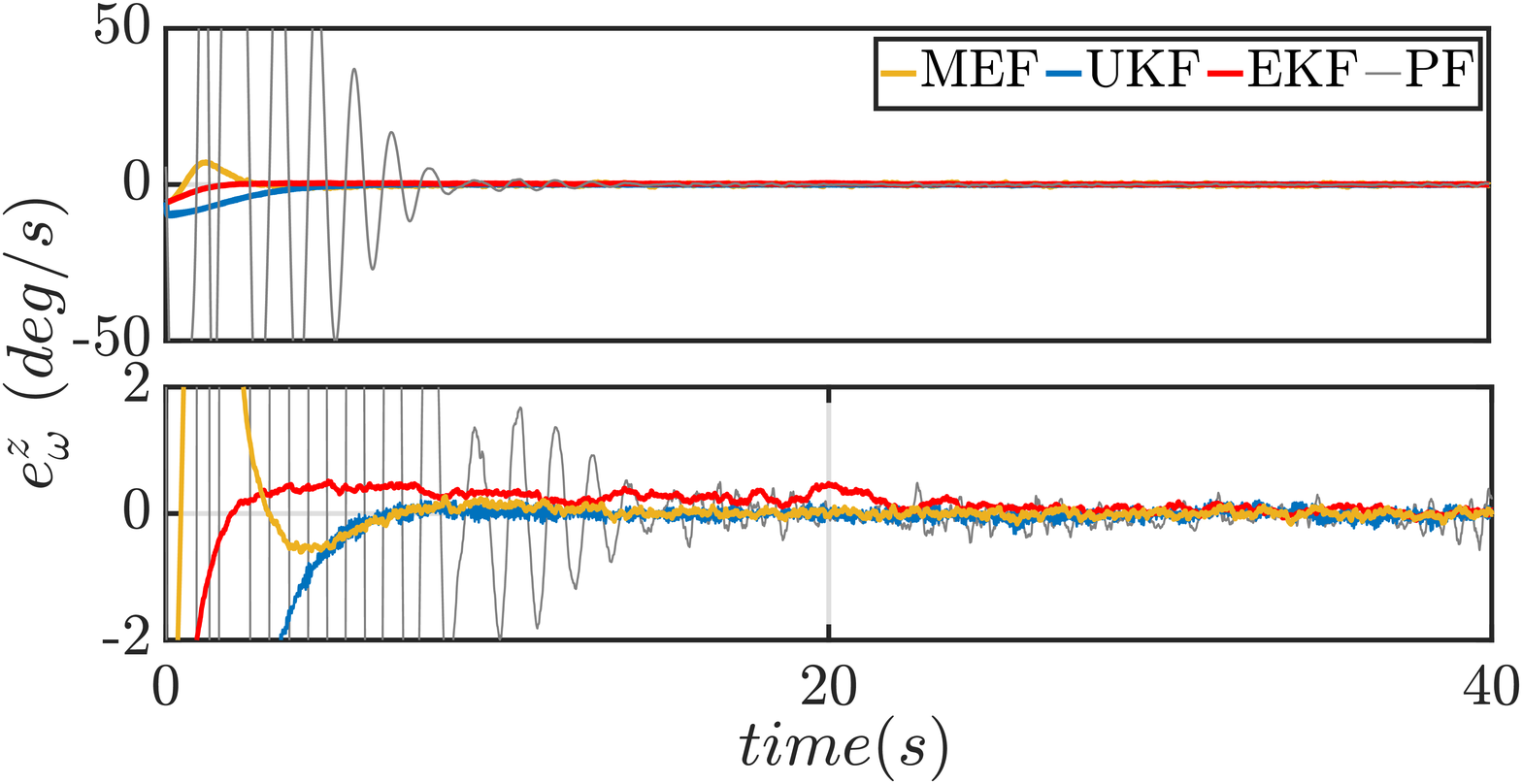}}
\end{tabular}
\end{center}
\caption{\small{Angular velocity estimation error (process noise-case 1). $\mathrm{(X, Y, Z)}$ component (left to right)}}
\label{fig1}
\end{figure}

In Fig. \ref{fig1}, the three components $\mathrm{(X, Y, Z)}$ of the angular rate estimation error are shown for each case, respectively. In the case of the MEF (depicted in yellow), the estimation errors converge after a very brief transient response providing with the smallest steady state error value of all filters. In the case of the PF (depicted in grey), $\Sigma=0.3\cdot10^{-3}$ was selected, which satisfies the covariance constraint imposed by (\ref{constr})-(\ref{eq36}), with $\mathrm{trace}(M)=0.63$. Note that the covariance constraint induces a trade-off regarding the PF's transient response, due to the fact that increased measurement noise levels require larger value for $\Sigma$; this results in slower transient response. As in the case of the MEF, the PF has also a short transient response but presents an oscillatory behavior in the steady state. This is attributed to the non-adaptive nature of the filter, and to the fact that measurement error in the optimal correction term $\d^*(t)$ is scaled by an almost constant matrix, as can be seen from \eqref{eq48}.

Regarding the stochastic filters, the EKF (depicted in red) seems to outperform the UKF (depicted in blue) as the former has a fast convergence, while the latter presents an oscillatory behavior.\par
\begin{figure}[h!]
\centering
{\includegraphics[scale=0.17]{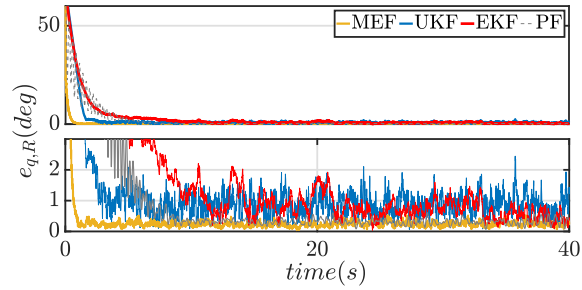}}
\caption{Attitude estimation error $e_{q,R}(t)$ (process noise-case 1)}
\label{fig2}
\end{figure}

\begin{figure}[h!]
\vspace{0.1cm}
\begin{tabular}{ccc}
\hspace{-0.39cm}
{\includegraphics[scale=0.12]{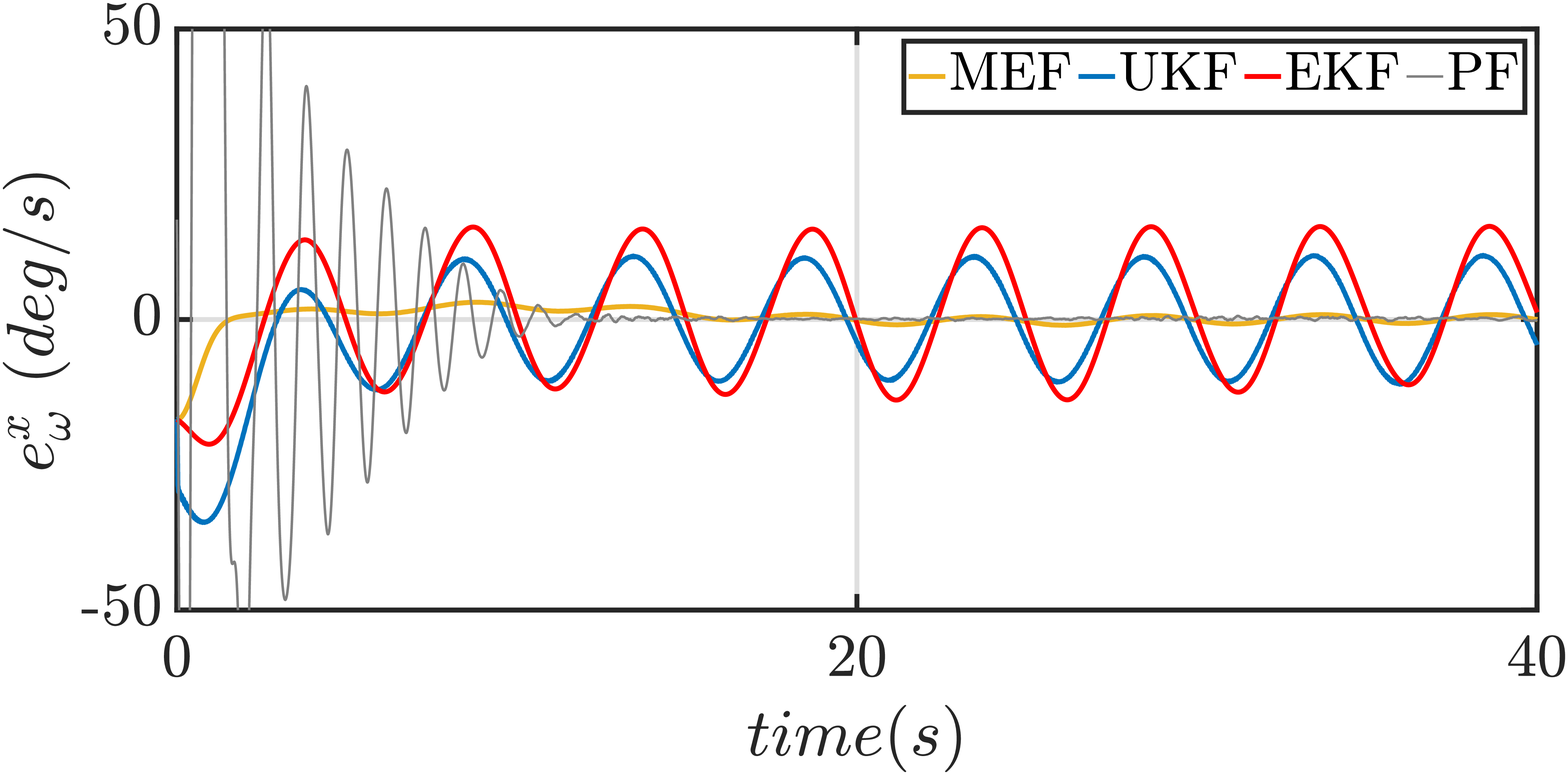}} \hspace{-0.8cm}
&{\includegraphics[scale=0.12]{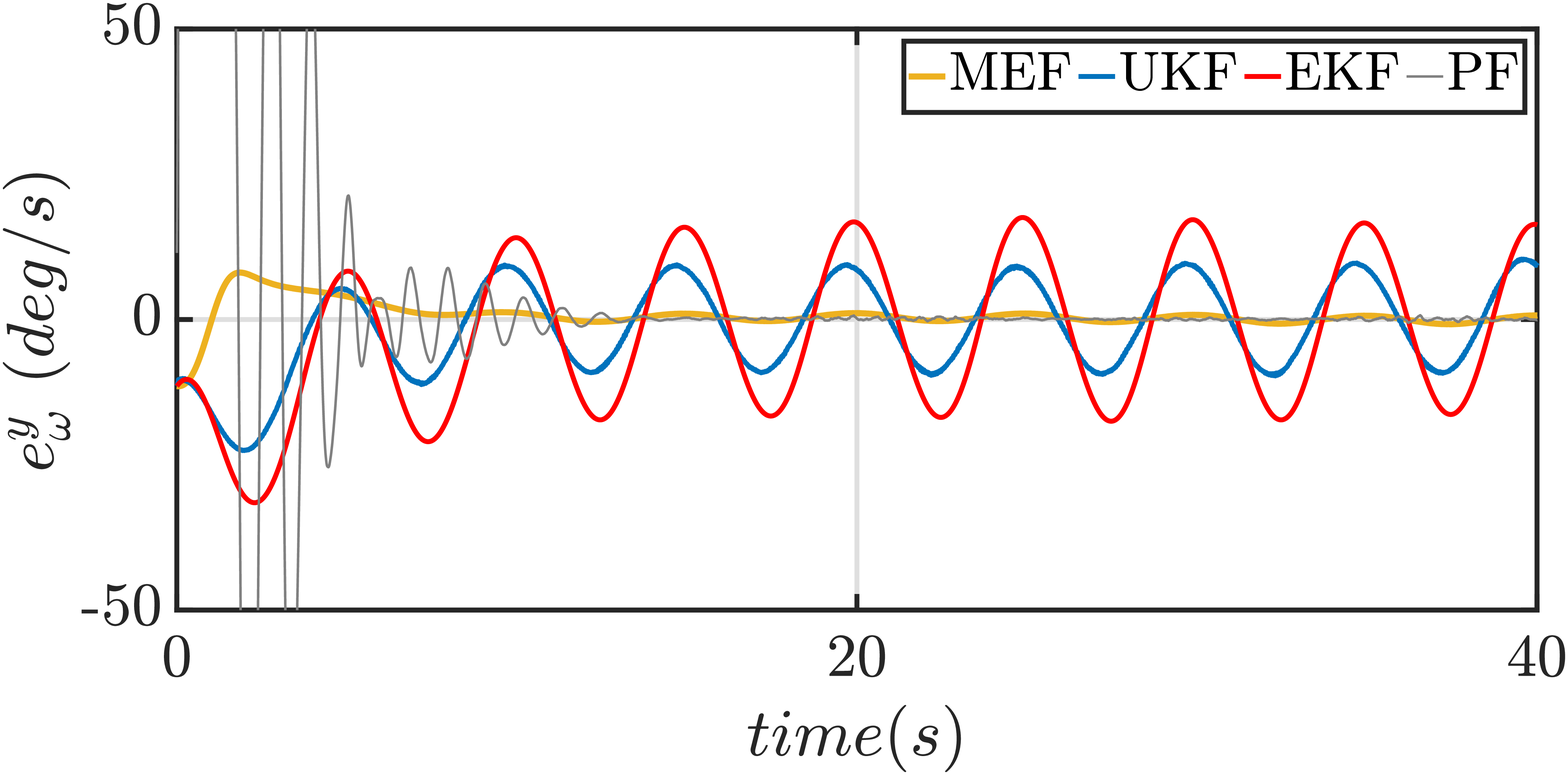}}\hspace{-0.8cm}
&{\includegraphics[scale=0.12]{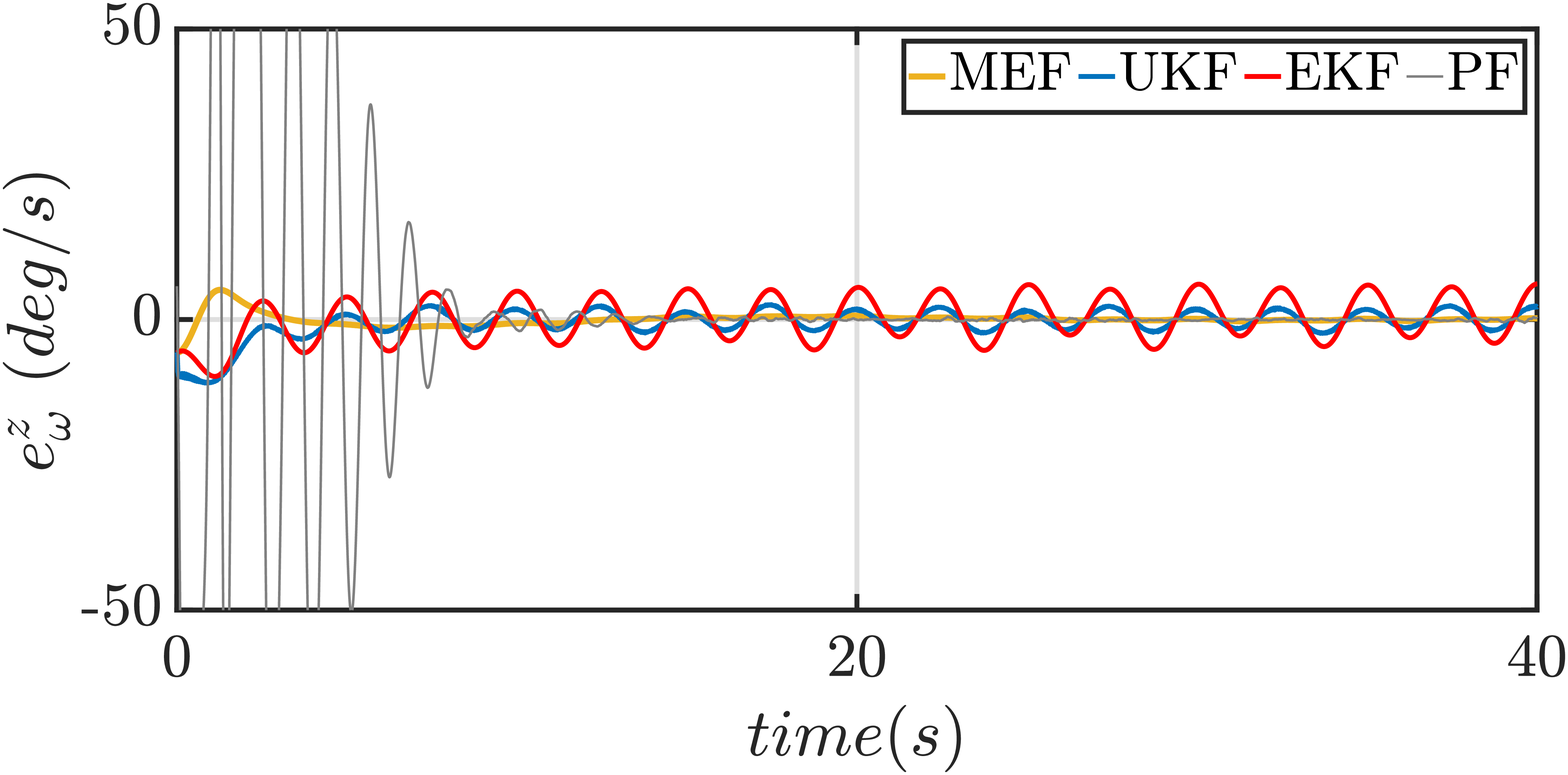}}
\end{tabular}
\caption{\small{Angular velocity estimation error (model error-case 1). $\mathrm{(X, Y, Z)}$ component (left to right)}}
\label{fig3}
\end{figure}

\begin{figure}
\centering
{\includegraphics[scale=0.15]{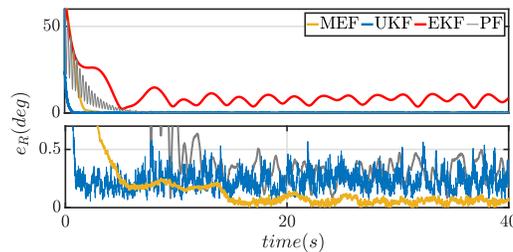}}
\vspace{-0.3cm}
 \caption{\small{Attitude estimation error $e_{q,R}(t)$ (model err
or-case 1)}}
\label{fig4}
\end{figure}

 This is because the rate estimates depend on the orientation estimates, which are re-projected many times within the algorithm. 
Another reason for the UKF's noisy asymptotic performance is the ad-hoc fine-tuning in our experiments. Nevertheless, such an approach is necessary to balance efficiency and extreme computational burden.
For the orientation error presented in Fig. \ref{fig2}, the MEF again shows its superiority by having the fastest transient response and smallest asymptotic error, while the predictive filter has a small angle bias due to the remaining angular velocity error. This is because the PF's orientation correction is made exclusively through the axis of rotation, and the kinematics remain isolated for geometric integration. However, the PF achieves the second-lowest estimation error with the lowest computational cost. The downside of the PF is that it needs precise tuning and many iterations in order for the estimates to be statistically consistent.\\ Thus, on the one hand, the PF architecture avoids an additional re-projection step and an expensive implementation; on the other hand, it leads to a constant deviation of around $0.056^o$ due to the lack of additional kinematic correction. 
Worth noticing, however, is that the kinematics express Euler's theorem and thus cannot be considered uncertain. The EKF converges smoothly towards zero, whereas the UKF appears to have an additional peak. This is because the rotation axis has not been estimated well up to that time step (Fig. \ref{fig1}). 
At the same time, the EKF outperforms the UKF during the steady-state. The reason is that the latter employs a stochastic linearization including mainly addition operations to produce the prediction and correction state, thus requiring intermediate normalization steps.
Another compelling observation is that for the PF, higher scaled gain can result in faster convergence at the cost of increasing the asymptotic estimation error.
Fig. \ref{fig3} depict the angular rate estimation in the case where a deterministic model error acts on the system dynamics. Both deterministic filters perform very well since they determine the necessary model error that drives the actual system and produces the obtained observations. In particular, by reducing the measurement noise, both the MEF's and the PF's rate estimation error converge fast to zero, and the same holds for the attitude estimation error (Fig. \ref{fig4}). For the PF however, the oscillatory behavior that has been observed previously, appears only within the transient state in this case.
\par
On the other hand, the stochastic filters' rate is affected significantly as the model error frequencies are transferred in the angular velocity error. In terms of the orientation error (Fig. \ref{fig4}), the deterministic model uncertainty influences only the EKF's response, while the UKF's remains unaffected, due to the stochastic linearization process. The deterministic model error is canceled out in the predicted state covariance step of the algorithm (proof in Appendix  \ref{AppendixA}). In particular, it is shown that there are sigma-point distributions that block the model error influence in the orientation estimate; hence, the model error appears only in the angular velocity's correction step. Thus, the UT -for particular sigma point distributions- potentially recasts the angle estimates uncontrollable from the model error. However, this does not mean that the UKF estimates the orientation correctly since the estimated rotation axis deviates significantly from its nominal trajectory; the filter remains blind w.r.t. model errors and trusts more its angle estimates. Additionally, the fact that the corrected angular velocity is a linear combination of the model error and the scaled output error, preserves the frequencies of all three components which appear unaltered in the angular velocity estimation error (Fig. \ref{fig3}).
The exact opposite is true in the EKF; both the rate and angle estimates are affected by the model uncertainty, as the Jacobian matrix (prediction update) is a function of the predicted rate estimate, and its components appear to all its entries. By utilizing the matrix inversion lemma for the output covariance matrix, it can be shown explicitly that the model error vector appears both in the upper and in the lower part of the gain matrix.
\subsubsection{Case 2: Attitude and rate estimation for a satellite}\label{Section5b}
\begin{figure}[h!]
\begin{adjustwidth}{-0.2cm}{}
\begin{center}
\vspace{0.1cm}
\begin{tabular}{ccc}
\hspace{-0.39cm}
{\includegraphics[scale=0.12]{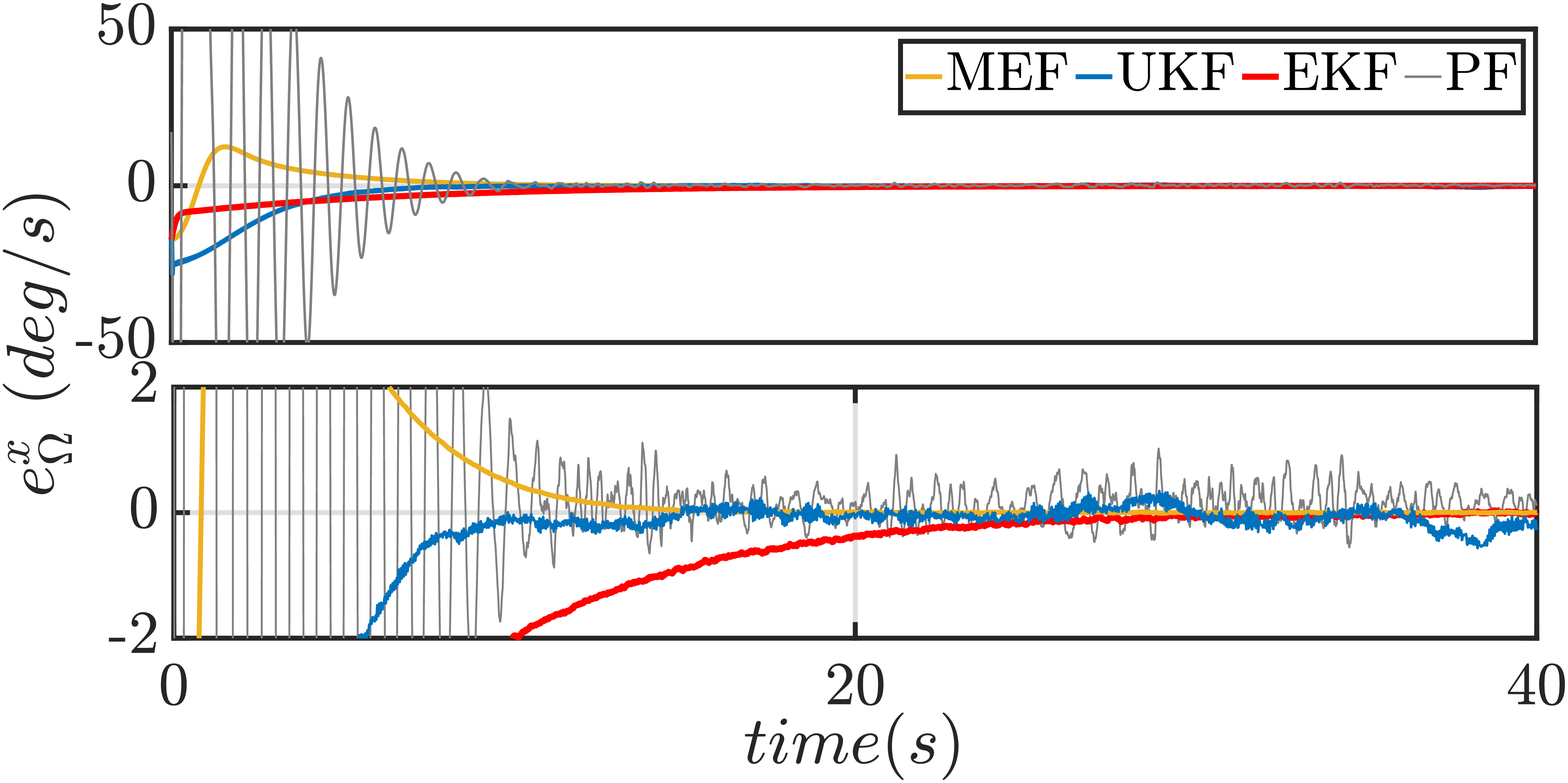}} \hspace{-0.8cm}
&{\includegraphics[scale=0.12]{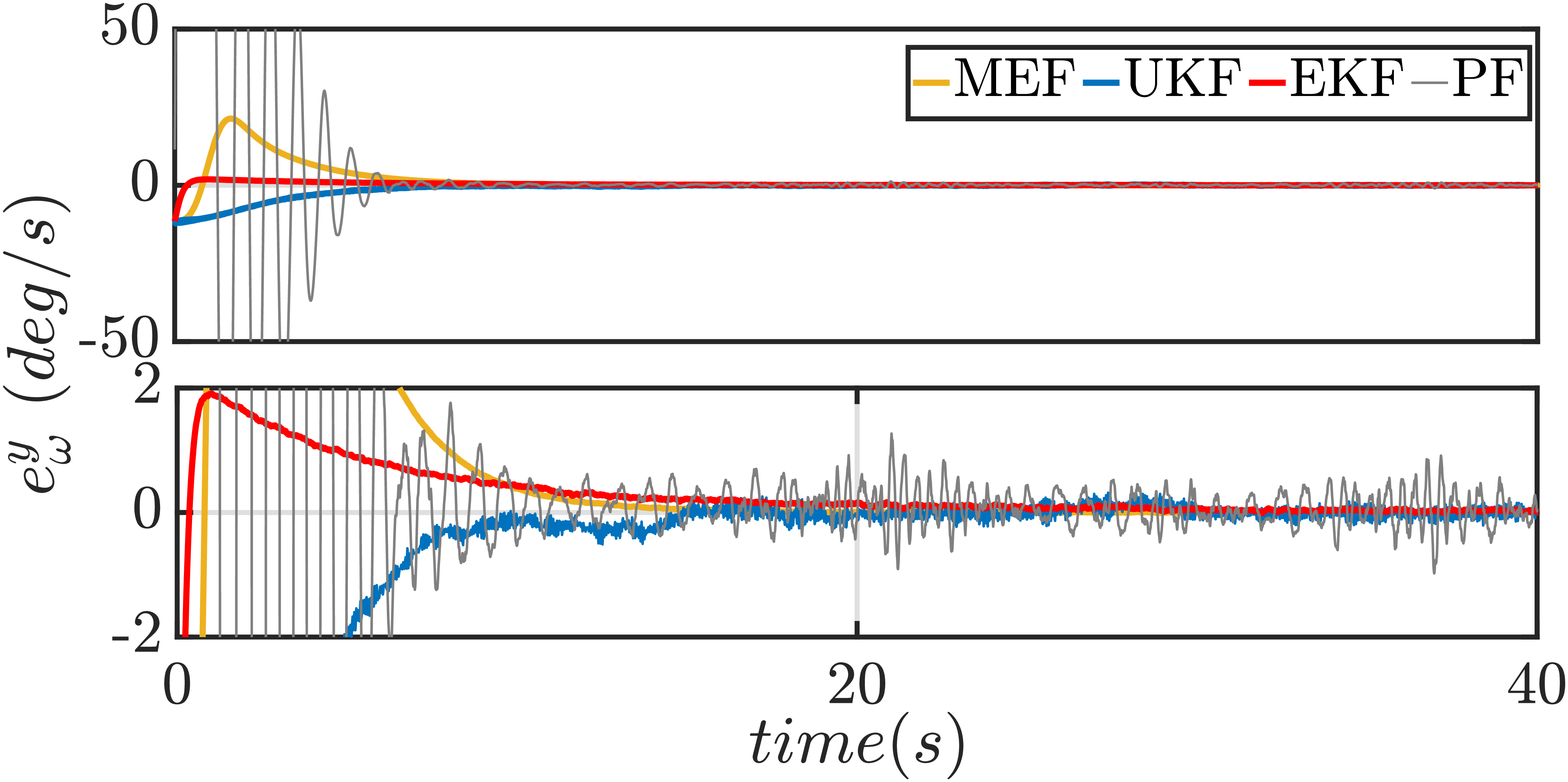}}\hspace{-0.8cm}
&{\includegraphics[scale=0.12]{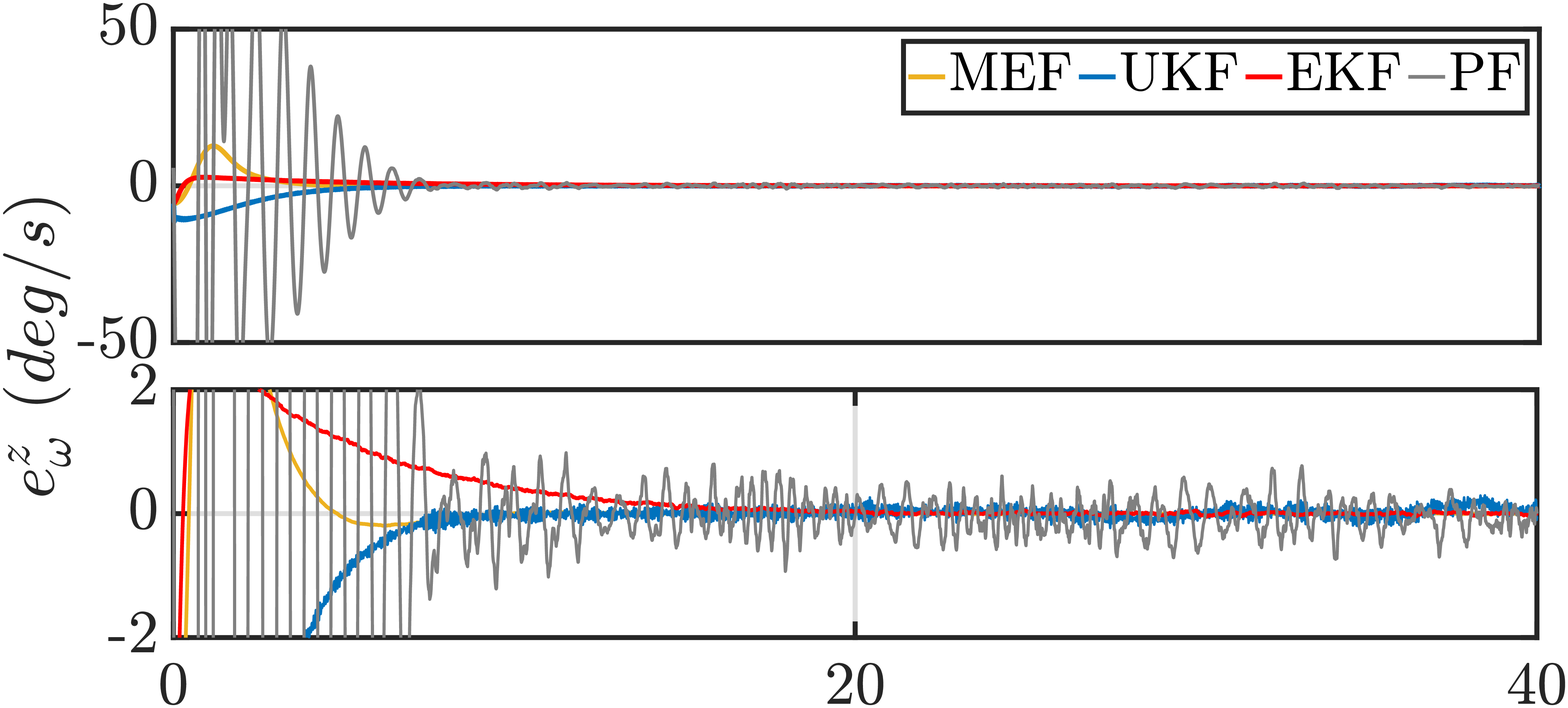}}
\end{tabular}
\end{center}
\caption{Angular velocity estimation error (process noise-case 2). X-Y-Z component (from top to bottom)}
\label{fig5}
\end{adjustwidth}
\end{figure}

\begin{figure}[h!]
\begin{center}
\begin{tabular}{c}
\hspace{-0.25cm}
{\includegraphics[scale=0.15]{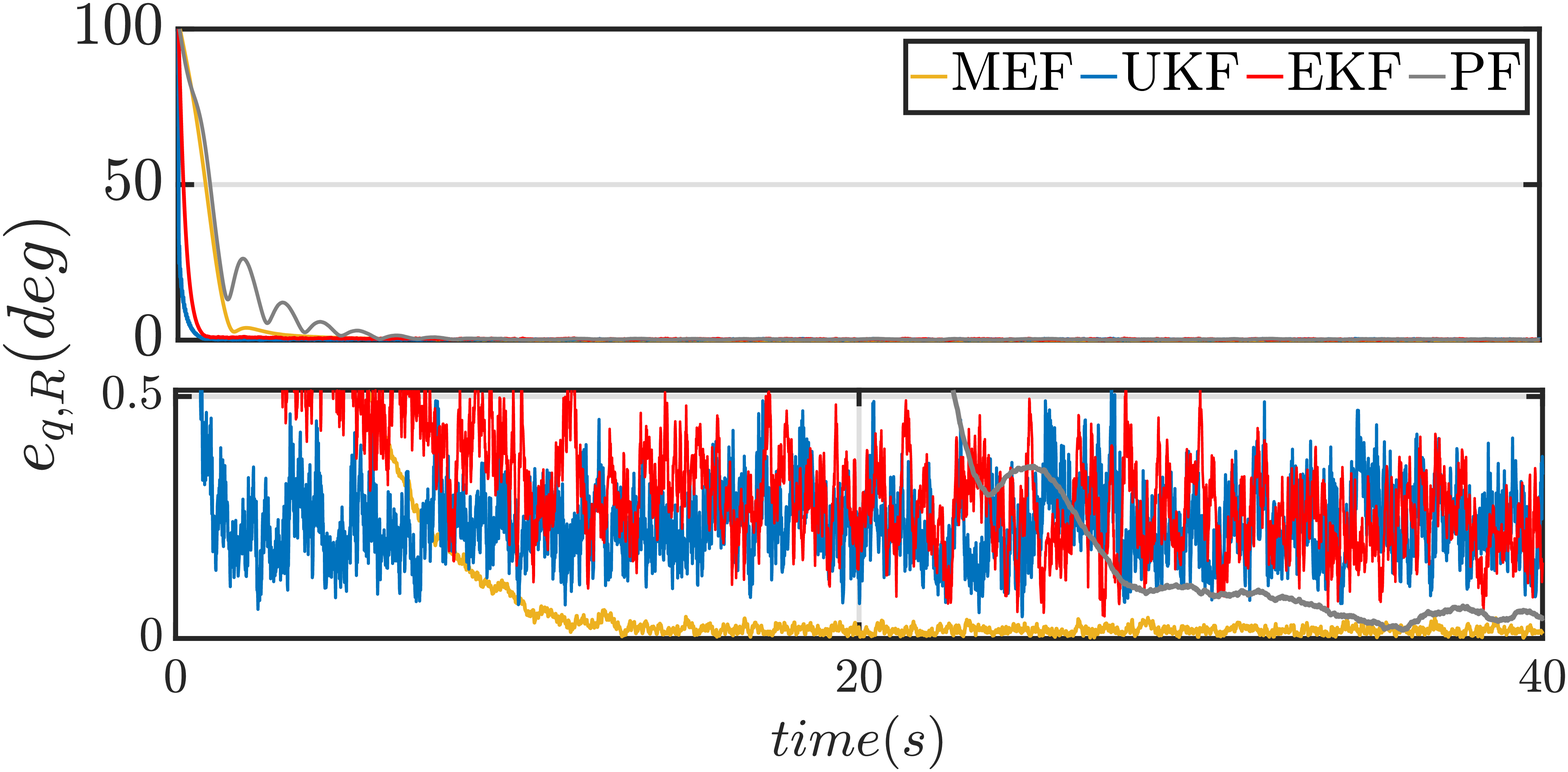}} \end{tabular} 
 \end{center}
 \caption{Attitude estimation error $e_{q,R}(t)$ (process noise-case 2)}
\label{fig6}
\end{figure}

\begin{figure}[h!]
\vspace{0.1cm}
\begin{tabular}{ccc}
\hspace{-0.39cm}
{\includegraphics[scale=0.12]{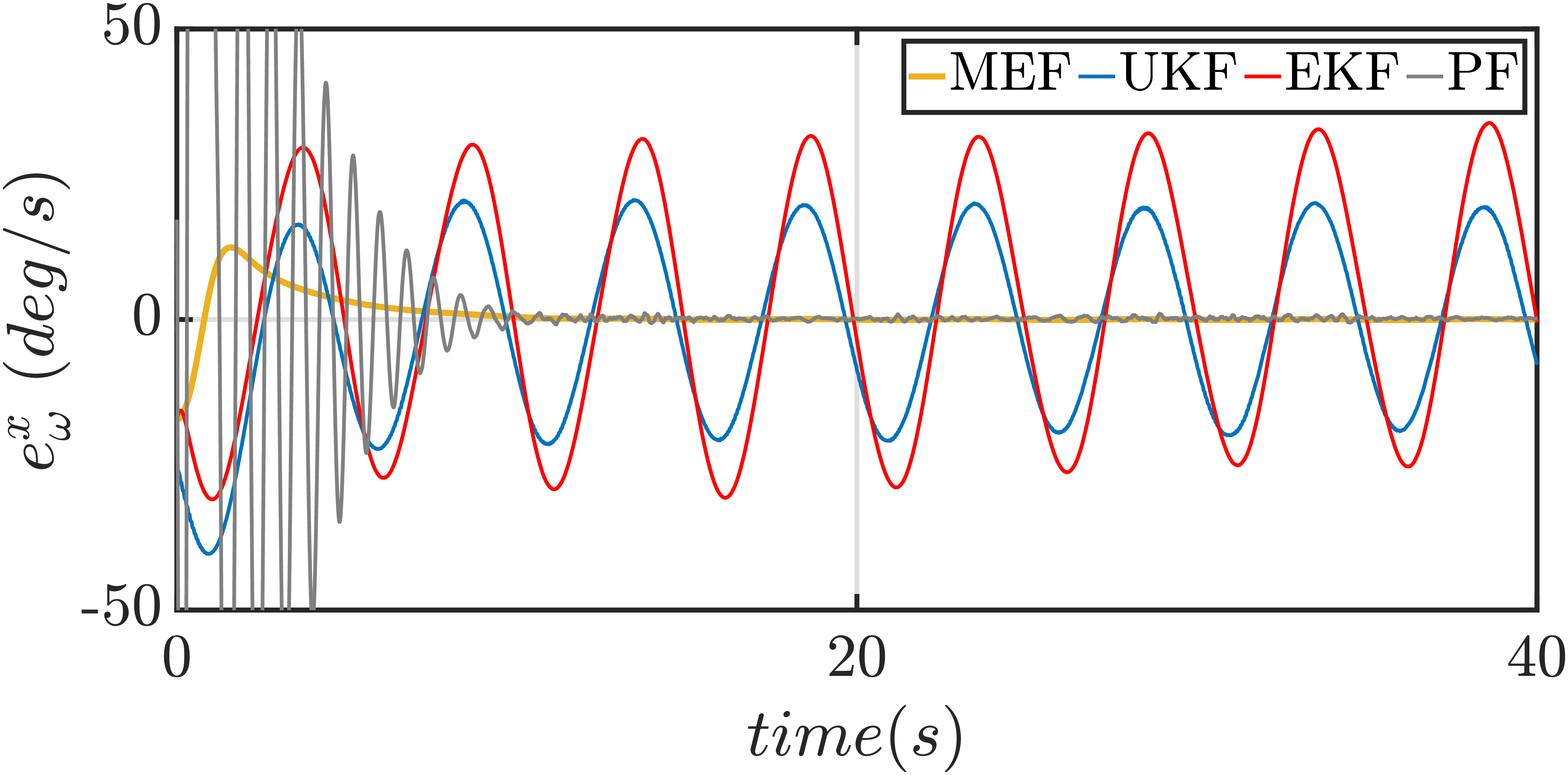}} \hspace{-0.8cm}
&{\includegraphics[scale=0.12]{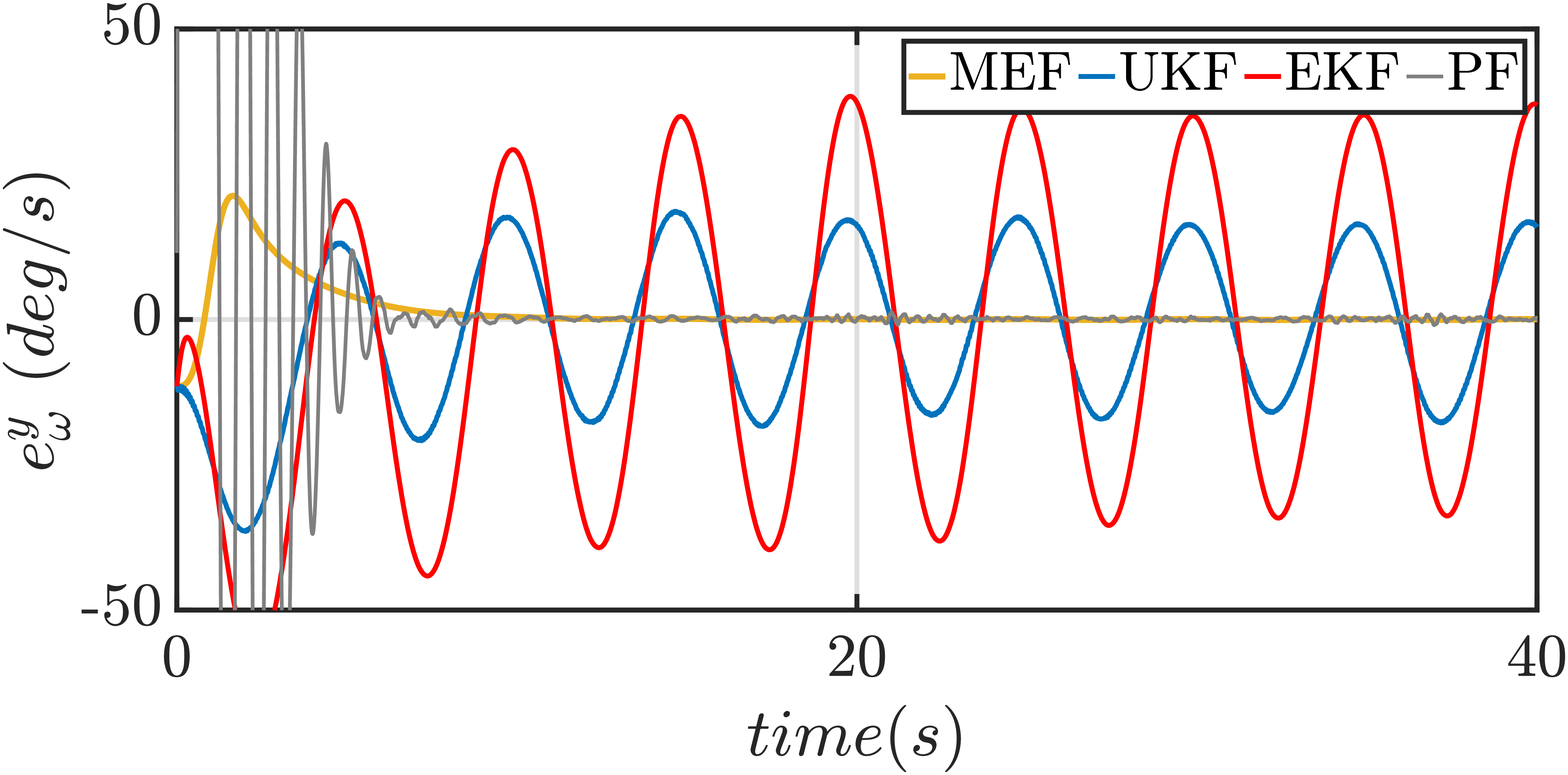}}\hspace{-0.8cm}
&{\includegraphics[scale=0.12]{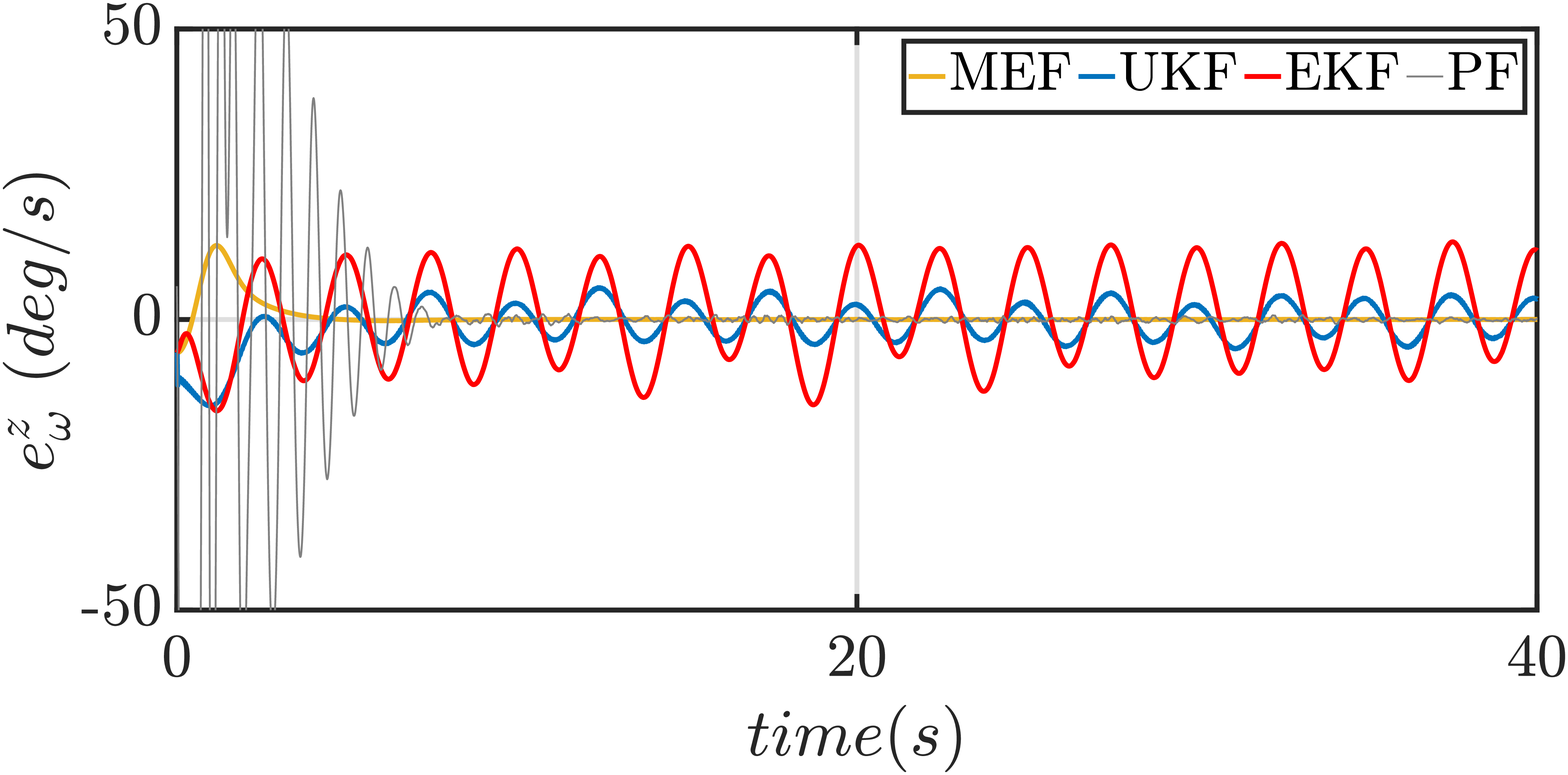}}
\end{tabular}
\caption{Angular velocity estimation error (model error-case 2). $\mathrm{(X, Y, Z)}$ component (left to right)}
\label{fig7}
\end{figure}

\begin{figure}[h!]
\begin{center}
\begin{tabular}{c}
\hspace{-0.19cm}
{\includegraphics[scale=0.17]{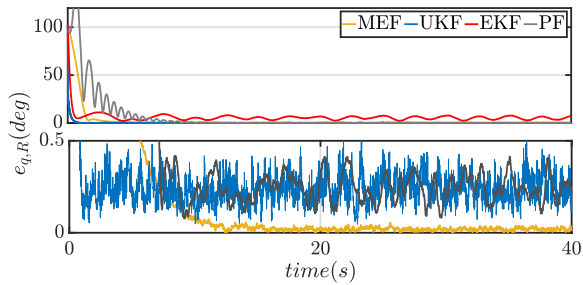}}
 \end{tabular}
 \end{center}
 \caption{Attitude estimation error $e_{q,R}(t)$ (deterministic model error -case 2)}
 \label{fig8}
\end{figure}

Fig. \ref{fig5}-\ref{fig8} show the performance of all the filters in the satellite study. Note that the input torques' low frequency and the increased moment of inertia result in a much slower attitude motion. Therefore, the filters converge faster towards zero both in attitude and rate. The MEF, once again, outperforms the rest of the filters by showcasing a similar behaviour as in the UAV case. 
Furthermore, the Gaussian approximate filters can be re-tuned to converge faster. In addition, the decreased measurement noise, allows us to opt $\Sigma=0.3\cdot10^{-4}$, for the PF and thus achieving a fastest transient and an improved asymptotic error.
\par
Regarding the model error case we observe once again that the deterministic filters outperform the stochastic ones, since both the axis of rotation and angle estimates present low transient and asymptotic error. The EKF as well as the UKF transfer the model error unhurt in the rate error as it is imposed by their architecture. Once again the UKF's orientation estimates are not affected by the model uncertainty for the same structural reasons mentioned previously.

\newpage
\section{Conclusions}\label{Section6}
This work performed a critical assessment of the reasons governing the superiority of deterministic modelling over stochastic, for the problem of orientation and rate estimation from vector measurements. The distinction between the two approaches was emphasised and investigated, with the state space's geometry and characteristics being the main criterion. By the analysis and the results of extensive simulations, the deterministic approach was shown to overcome important deficiencies imposed by the Bayesian architectures, and to handle large model errors. As an example, the second-order-optimal Minimum Energy Filter (MEF) \cite{saccon} was presented, and a modified predictive filter (PF) on the $\mathbb{TSO}(3)$ was derived. Both of these filters were compared versus the most commonly used representatives of the Gaussian Approximate Filters, the EKF and the UKF. 
Two different simulation cases were considered, for a UAV and for a satellite, respectively. The simulations revealed that the deterministic filters, and in particular  the MEF, outperform the Gaussian approximate solutions especially in the realistic scenario, where a deterministic model error exerts on the actual plant. The reason is fundamental -from first principles- and originates in the set-theoretic approach for estimation, when seen as a dual optimal control problem. The stochastic filters require at least one re-projection step and are affected by model errors. In particular, we address that quaternion normalization leads to unbiasedness of the orientation and rate estimates. In addition, for certain sigma point distributions, the UKF's estimation angle is uncontrollable from model errors.
While a more efficient implementation of the UKF for attitude estimation exists~\cite{cheon2007unscented} where the stochastic linearization is performed by utilizing intrinsic gradient descent algorithms, it is not robust w.r.t. deterministic model errors and also requires one re-projection step.\par
Another remark is that both stochastic filters require the initial prior information in contrast with the deterministic ones. In practice this information may not be available. For example, satellite missions are placed in environments that are not fully known beforehand, which makes it impossible to obtain data in advance. Deterministic filters do not require any prior initialization, providing exceptional flexibility for the problem at hand. From the deterministic filters presented in this work, the predictive filter has to be tuned to provide statistically consistent results. However, this tuning is based on the measurement noise statistics, which can be determined offline by experimentation. Having the disadvantage of an almost fixed gain, the predictive filter is still to be investigated as future research under adapted gain scaling. The EKF and UKF have roughly the same accuracy. Thus, due to  the computational overhead of the UKF, the simplicity of the Jacobian matrix calculations, and the quasi-linear nature of the quaternion kinematics the EKF is considered preferable compared to the UKF for the task. Per contra, both deterministic filters -and especially the MEF- perform better as they achieve lower errors in both cases. Finally,  
a further analysis for examining the filters' limitations is a potential future research objective. Such operating factors include eclipse conditions, event-triggered change of dynamics \cite{hu2020event}, co-linearity of measurements, extreme measurement noise, risk-averse events \cite{kalogerias2020better}, etc.

\section{Appendix}\label{Appendix}

\subsection {Intrinsic Lie derivatives for predictive filter}\label{AppendixA}
The predictive filter on $\mathbb{TSO}(3)$ requires knowledge of the terms $\zita_{k}(\widehat{R},\widehat{\oo},h;t)$ and $w_k(\widehat{R},\widehat{\oo},h)$. It is 
\begin{equation}\label{eq50}
\begin{aligned} 
\zita_{k}\left(\widehat{R}, \widehat{\oo}^{\times}, h;t\right)&= h\mathcal{L}_{\f}^{1}(\widehat{\y}_k)+\frac{h^2}{2!}\mathcal{L}_{\f}^{2}(\widehat{\y}_k) 
\end{aligned}
\end{equation}
By defining the inverse map $(~)^{-\times}: \mathfrak{so}{(3)} \rightarrow \mathbb{R}^{3}$, the system can be written as:
\begin{equation}\label{eq51}
\begin{aligned} 
(\widehat{R}^{\top} \dot{\widehat{R}} )^{-\times}&=\widehat{\oo} \\
 \dot{\widehat{\oo}} &=\mathbb{I}^{-1}\left((\mathbb{I}\widehat{\oo})^{\times} \widehat{\oo}+\boldsymbol{T}\right)+G\d~.\\ 
 \end{aligned}
 \end{equation}
Furthermore, by declaring
\begin{equation}
\begin{aligned}
\f(\widehat{\oo})=\left[\begin{array}{c}  {\widehat{\oo}} \\ {\mathbb{I}^{-1}\left((\mathbb{I}\widehat{\oo})^{\times} \widehat{\oo}+\boldsymbol{T}\right)+G \d} \end{array}\right],
\end{aligned}
\end{equation}
with $\widehat{\oo} \in \mathbb{R}^3$, the first term of  \eqref{eq50} reads:
\begin{equation}
\begin{aligned}
\mathcal{L}_{\f}^{1}(\widehat{\y}_k)\\
&=\big({\partial}_{_{(\widehat{R},\hat{\oo})}}{\widehat{\y}_k}\big)\f\\
&=\big({\partial}_{_{(\widehat{R},\hat{\oo})}}{\widehat{R}^{\top}\a_k}\big)\f\\
&= \left[\begin{array}{cc} {\partial}_{\widehat{R}}{\widehat{R}^{\top}\a_k} & {\partial}_{\widehat{\oo}}{\widehat{R}^{\top}\a_k} \end{array} \right]\cdot
 \left[\begin{array}{c} {\widehat{\oo}} \\ {\mathbb{I}^{-1}\left((\mathbb{I}\widehat{\oo})^{\times} \widehat{\oo}+\boldsymbol{T}\right)+G \d} \end{array}\right] \\
 \end{aligned}
\end{equation}
In order to calculate the first term in the brackett, we consider a deviation $\widehat{\d \R} $ from $\widehat{R}$ with $\widehat{\d \boldsymbol{R}} =\text{exp}(\widehat{\oo}^{\times})$ and $\widehat{\oo}$ a tangent vector attached on the identity. Then
\begin{equation}
\begin{aligned}
\mathbf{\partial}_{\widehat{R}}{\widehat{R}^{\top}\a_i} \\
&=\mathbf{\partial}_{\widehat{R}}({\widehat{R}^{\top}\a_i}) (\widehat{\d \boldsymbol{R}})\\
&= \mathbf{\partial}_{\hat{R}}({\hat{R}\widehat{\d \boldsymbol{R}})^{\top}\a_i}\\
&={\partial_{\widehat{\oo}}}\big(e^{-\widehat{\oo}^{\times}}{\widehat{R}^{\top}\a_i}\big)|_{\widehat{\oo}=0} \\
&=\big({\widehat{R}^{\top}\a_i}\big)^{\times} \\
&= \widehat{\y}_i^{\times}
\end{aligned}
\end{equation}
\newline

The latter results using the Taylor expansion of the exponential matrix, and from the fact that with $X=\x^{\times} \in \mathfrak{so}(3)$
\begin{equation}
\begin{aligned}
 \mathbf{\partial}_{X}\big({X^{\top}\a}\big) &={\partial}_{x}\big(({\x^{\times})^{\top}\a}\big) \\
 &={\partial}_{\x}\big(-\x^{\times}\a\big)\\
 &={\partial}_{\x}\big(\a{\times}\x\big)\\
  &={\partial}_{\x}\big(\a^{\times}\x\big)\\
 &=\a^{\times} \in \mathfrak{so}(3)
\end{aligned}
\end{equation}

Finally, since $ \mathbf{\partial}_{\widehat{\oo}}({\hat{R}^{\top}\a_k}) =0$ we obtain
\begin{equation}
\begin{aligned}
\mathcal{L}_{\f}^{1}(\widehat{\y}_k)&= \big({\widehat{R}^{\top}\a_k}\big)^{\times} \widehat{\oo}=- \widehat{\oo}^{\times}\big({\widehat{R}^{\top}\a_k}\big) + 0
\end{aligned}
\end{equation}
Furthermore, the second-order Lie derivative of \eqref{eq50} reads
\begin{equation}
\begin{aligned}
\mathcal{L}_{\phi}^{2}(\hat{\y}_i)&=\mathcal{L}^{1}\big(- \widehat{\oo}^{\times}\big({\widehat{R}^{\top}\a_k}\big)\big) \\
&=- \left[\begin{array}{cc} \mathbf{\partial}_{\widehat{R}}{\big(\widehat{\oo}^{\times}\big({\widehat{R}^{\top}\a_k}\big)\big)} & \mathbf{\partial}_{\widehat{\oo}}\big({\widehat{\oo}^{\times}\big({\widehat{R}^{\top}\a_i}\big)\big)} \end{array} \right] \cdot\\
&\left[\begin{array}{c}  {\widehat{\oo}} \\ {\mathbb{I}^{-1}\left((\mathbb{I}\widehat{\oo} )^{\times} \widehat{\oo}+\boldsymbol{T}\right)+G \d} \end{array}\right] \\
&=- \left[\begin{array}{cc} {\widehat{\oo}^{\times}\big({\widehat{R}^{\top}\a_i}\big)^{\times}} & -{\big({\widehat{R}^{\top}\a_k}\big)^{\times}} \end{array} \right] \cdot \\
&\left[\begin{array}{c}  {\widehat{\oo}} \\ {\mathbb{I}^{-1}\left((\mathbb{I}\widehat{\oo})^{\times} \widehat{\oo}+\boldsymbol{T}\right)+G \d} \end{array}\right] \\
&=\big(\widehat{\oo}^{\times}\big)^{2}\big({\widehat{R}^{\top}\a_k}\big)+\\&{\big({\widehat{R}^{\top}\a_k}\big)^{\times}} \left({\mathbb{I}^{-1}\left((\mathbb{I}\widehat{\oo})^{\times} \widehat{\oo}+\boldsymbol{T}\right)+G \d}\right)
 \end{aligned}
\end{equation}

where the term $\mathbf{\partial}_{\hat{R}}{\big(\hat{\oo}^{\times}\big({\hat{R}^{\top}\a_k}\big)\big)}$ is computed using the product rule:

\begin{equation}
\begin{aligned}
\mathbf{\partial}_{\widehat{R}}{\big(\widehat{\oo}^{\times}\big({\widehat{R}^{\top}a_k}\big)\big)} &=  \widehat{\oo}^{\times}{\partial}_{\widehat{\oo}}\big({\hat{R}^{\top}\a_i}\big)|_{\widehat{\oo}=0}\\
&= \widehat{\oo}^{\times}\Big({\partial}_{\widehat{\oo}}\big( e^{-\widehat{\oo}^{\times}}\big){\widehat{R}^{\top}\a_k}\Big)|_{\widehat{\oo}=0}\\
&= \widehat{\oo}^{\times} \big({\widehat{R}^{\top}\a_k} \big)^{\times}
\end{aligned}
\end{equation}

\subsection {UKF with deterministic model error}\label{AppendixB}

Let us declare with $\uu$ the nominal input to the filter. Then, we can write $\tilde{\uu}=u+\mathbb{I}\d$ that is, the input torques corrupted by the model error $\d$ as they applied to the model. The filter utilises the equations 

\begin{equation}
\begin{aligned}
d{\x}=\f(\x,\uu)=\f(\x,\tilde{\uu})-\tilde{G}\mathbb{I}\d
\end{aligned}
\end{equation}
where the first $n_1$ rows of $\tilde{G}$ refer to the kinematics and therefore are zero.
The predicted sigma-points are then calculated according to 

\begin{equation}
\begin{aligned}
{\x}_{k \mid k-1}^{\d}&={\f}\left({\x}_{k-1}, \tilde{\uu}_{k-1}\right)-G\mathbb{I}\d_{k-1}\\
&={\x}_{k \mid k-1}-G\mathbb{I}\d_{k-1}
\end{aligned}
\end{equation}

and the predicted state according to 

\begin{equation}
\begin{aligned}
\hat{{\x}}_{k}^{\d}=&\sum_{i=0}^{2 L} w_{i}^{(m)} {\x}_{i, k \mid k-1}^{\d}\\
&=\sum_{i=0}^{2 L} w_{i}^{(m)}{\f}\left({\x}_{i,k-1}, \tilde{\uu}_{k-1}\right)-\sum_{i=0}^{2 L} w_{i}^{(m)}G\mathbb{I}\d_{k-1}\\
&=\sum_{i=0}^{2 L} w_{i}^{(m)}{\f}\left({\x}_{i,k-1}, \tilde{\uu}_{k-1}\right)-G\mathbb{I}\d_{k-1}\sum_{i=0}^{2 L} w_{i}^{(m)}
\end{aligned}
\end{equation}
By utilizing the scaled UT

\begin{equation}
\begin{aligned}
\sum_{i=0}^{2 L} w_{i}^{(m)}=\frac{2L}{2(L+\lambda)}
\end{aligned}
\end{equation}
where $\lambda=\alpha^2(L+k)-L$ \cite{van2004sigma}. From  here we observe that for $k=0$ and $\alpha=1$ 

\begin{equation}
\begin{aligned}
\hat{{\x}}_{k}^{\delta}
&=\sum_{i=0}^{2 L} w_{i}^{(m)}{\f}\left({\x}_{i,k-1}, \tilde{\uu}_{k-1}\right)-G\mathbb{I}\d_{k-1}\\
&=\hat{\x}_k- G\mathbb{I}\d_{k-1}
\end{aligned}
\end{equation}
Therefore, the predicted state covariance 

\begin{equation}
\begin{aligned}
{P}_{{\x}_{k}}^{\d}&=\sum_{i=0}^{2 L} w_{i}^{(c)}\left({\x}_{i, k \mid k-1}^{\d}-\hat{{\x}}_{k}^{\d}\right)\left({\x}_{i, k \mid k-1}^{\d}-\hat{{\x}}_{k}^{\d}\right)^{\top}\\
&=\sum_{i=0}^{2 L} w_{i}^{(c)}\left({\x}_{i, k \mid k-1}^{}-\hat{{\x}}_{k}^{}\right)\left({\x}_{i, k \mid k-1}^{}-\hat{{\x}}_{k}^{}\right)^{\top}\\
&={P}_{{\x}_{k}}
\end{aligned}
\end{equation}
The same applies to the cross covariance matrix where the model error term is canceled out. Thus, the adaptive gain of the filter is not affected by the model error. The only step where the error applies is the correction step through the rate-part of $\widehat{\x}^{*}_k$.
\subsection {Bias due to quaternion re-projection}\label{AppendixC}
\textcolor{black}{The last step for both the EKF and UKF is a re-projection of the corrected quaternion }
\begin{equation}
\begin{aligned}
\widehat{q}_{k|k}=\widehat{q}_{k|k-1}+K_{u} {\e}_{k|k-1}
\end{aligned}
\end{equation}
where $K_u=K_{[1:4,1:6]}$ and $\epsilon_{k|k-1}=y_{k|k-1}-\widehat{y}_{k|k-1}$. The normalised corrected quaternion is given by

\begin{equation}\label{eq70}
\begin{aligned}
&\widehat{q}^{~*}_{k|k}=\frac{\widehat{q}_{k|k}}{||\widehat{q}^{}_{k|k}||}=
\frac{\widehat{q}_{k|k-1}+K_{u}\epsilon_{k|k-1}}{||\widehat{q}_{k|k-1}+K_{u}(\epsilon_{k|k-1})||}
\end{aligned}
\end{equation}
We are interested to examine the function $d:\mathbb{R}^4 \rightarrow \mathbb{R}$ with $d(X)=||X||^{-1}$ in a neighborhood of $\widehat{q}_{k|k} \in \mathbb{S}^3 \subset \mathbb{R}^4$ in the direction of $K_u \epsilon_{k|k-1}$.\\

We have 

\begin{equation}\label{eq71}
\begin{aligned}
&d(\widehat{q}_{k|k-1}+K_{u}\epsilon_{k|k-1})=\\
&d(\widehat{q}_{k|k-1})+\langle {\nabla{d}(X)}\big|_{X=\widehat{q}_{k|k-1}}, K_{u}\epsilon_{k|k-1} \rangle+\\
&\langle K_{u}\epsilon_{k|k-1}, \mathbb{H}(d)\Big|_{\widehat{q}_{k|k-1}}K_{u}\epsilon_{k|k-1}\rangle +...
\end{aligned}
\end{equation}
where $\mathbb{H}: \mathbb{R}^4\rightarrow \mathbb{R}^{4\times4}$ is the Hessian of $d$.
Given that $\widehat{q}_{k|k-1} \in \mathbb{S}^3$,  $d(\widehat{q}_{k|k-1})=1$.\par
\textcolor{black}{Furthermore}, $\nabla{d}{(X)}=\nabla{||X||}^{-1}=-\nabla{||X||}=||X||^{-1}X^{\top}$ and 

\begin{equation}\label{eq72}
\begin{aligned}
\mathbb[\mathbb{H}(d)]_{i,j}= \left\{
\begin{array}{ll}
     \frac{||X||^2-X^2_{i}}{||X||^3}&,~ i=j\\ \\

      -\frac{X_{i}X_{j}}{||X||^3}&,~ i\neq j\\
\end{array} 
\right. 
\end{aligned}
\end{equation}

\textcolor{black}{Thus, by ignoring the second order terms, \eqref{eq70} can be written as} 

\begin{equation}\label{eq73}
\begin{aligned}
&\widehat{q}^{~*}_{k|k}=\\
&(\widehat{q}_{k|k-1}+K_{u}\epsilon_{k|k-1})-\\
&(\widehat{q}_{k|k-1}+K_{u}\epsilon_{k|k-1})(\widehat{q}^{\top}_{k|k-1}K_u\epsilon_{k|k-1})+
%&[(\widehat{q}_{k|k-1}+K_u\epsilon_{k|k-1})^{{\top}} \mathbb{H}(d)(\widehat{q}^{}_{k|k-1}+K_u\epsilon_{k|k-1})]\cdot\\
%&~~~~~~~~~~~~~~~~~~~~~~~~~~~~~~~~~~~~~~~~~~~~~~~~~~~~~~~~~~~~~~~~~~(\widehat{q}^{\top}_{k-1|k}K_u\epsilon_{k|k-1})+
...
\end{aligned}
\end{equation}
\textcolor{black}{The latter equation shows the effect of the normalization step on the  (unbiased) corrected estimate of the EKF and UKF algorithm. By re-projecting the state on the unit sphere, a bias is induced that is propagated forward in time in the next iteration of the algorithm. This justifies the bias that appears in the figures of the orientation error for both the EKF and UKF filters.}

\bibliographystyle{plain}
\bibliography{refs}
\end{document}